\documentclass[traditabstract]{aa}
\usepackage{graphicx}
\usepackage{txfonts}
\usepackage{natbib}
\usepackage{lscape}
\usepackage{rotating} 
\bibpunct{(}{)}{;}{a}{}{,} 
\newcommand{\Msun}{\mbox{$\mathrm{M}_{\odot}$}}

\newcommand{\Teff}{\mbox{$T_{\mathrm{eff}}$}}

\newcommand{\Lines}[3]{\Ion{#1}{#2}\,$\lambda\lambda$\,#3}

\newcommand{\be}{\begin{equation}}
\newcommand{\ee}{\end{equation}}
\newcommand{\bea}{\begin{eqnarray}}
\newcommand{\eea}{\end{eqnarray}}
\newcommand{\bsube}{\begin{subequations}}
\newcommand{\esube}{\end{subequations}}

\newcommand{\Ion}[2]{#1{\,\scriptsize #2}}

\begin{document}
\title{White  dwarf-main  sequence  binaries   from  LAMOST:\\  the  DR1
  catalogue}

\author{
J.~J. Ren\inst{1,2},
A. Rebassa-Mansergas\inst{3},
A.~L. Luo\inst{1,2},
Y.~H. Zhao\inst{1,2},
M.~S. Xiang\inst{4},
X.~W. Liu\inst{3,4},
G. Zhao\inst{1,2},\\
G. Jin\inst{5},
Y. Zhang\inst{6}
}

\authorrunning{J. Ren et al.}
\titlerunning{WDMS binaries from LAMOST}

\institute{
Key Laboratory of Optical Astronomy, National Astronomical Observatories,
Chinese Academy of Sciences, Beijing, 100012, China\\
\email{jjren@nao.cas.cn, lal@nao.cas.cn}
\and
University of Chinese Academy of Sciences, Beijing, 100049, China
\and
Kavli Institute for Astronomy and Astrophysics, Peking University, Beijing, 100871, China
\and
Peking University, Beijing, 100871, China
\and
University of Science and Technology of China, Hefei, 230026, China
\and
Nanjing Institute of Astronomical Optics \& Technology, National Astronomical Observatories, Chinese Academy of Sciences, Nanjing, 210042, China
}

\date{Received ; accepted }

\abstract 
{White dwarf-main sequence  (WDMS) binaries are used  to study several
  different important open problems in modern astrophysics.}
{The Sloan Digital Sky Survey  (SDSS) identified the largest catalogue
  of WDMS binaries currently known.  However, this sample is seriously
  affected  by  selection  effects   and  the  population  of  systems
  containing   cool  white   dwarfs  and   early-type  companions   is
  under-represented.  Here  we search  for  WDMS  binaries within  the
  spectroscopic  data  release  1  of   the  LAMOST  (Large  sky  Area
  Multi-Object fiber Spectroscopic Telescope) survey.  LAMOST and SDSS
  follow  different target  selection algorithms.  Hence, LAMOST  WDMS
  binaries may  be drawn from  a different parent population  and thus
  help in overcoming the selection effects incorporated by SDSS on the
  current observed population.}
{We  develop  a  fast  and  efficient routine  based  on  the  wavelet
  transform to  identify LAMOST  WDMS binaries  containing a  DA white
  dwarf and  a M  dwarf companion,  and apply  a decomposition/fitting
  routine  to their  LAMOST spectra  to estimate  their distances  and
  measure their  stellar parameters, namely the  white dwarf effective
  temperatures, surface  gravities and masses, and  the secondary star
  spectral types.}
{We  identify  121   LAMOST  WDMS  binaries,  80  of   which  are  new
  discoveries,  and  estimate  the  sample to  be  $\sim$90\,per cent 
  complete.  The LAMOST   and  SDSS  WDMS  binaries are  found  to  be
  statistically different.   However, this  result is  not due  to the
  different   target  selection   criteria   of   both  surveys,   but
  likely  a simple  consequence  of  the different  observing
  conditions.  Thus,  the LAMOST  population is found  at considerably
  shorter  distances ($\sim$50--450\,pc)  and is  dominated by  systems
  containing early-type companions and hot white dwarfs.  }
{Even  though WDMS  binaries  containing cool  white  dwarfs are  also
  missed by the LAMOST survey, the LAMOST WDMS binary sample dominated
  by systems containing early-type companions is an important addition
  to  the  current  known   spectroscopic  catalogue.   Future  LAMOST
  observations however  are required to  increase the small  number of
  LAMOST WDMS binaries.}


\maketitle

\section{Introduction}

White dwarf-main sequence (WDMS)  binaries are detached compact binary
stars formed by a white dwarf  and a main sequence star. These binaries
evolve from main  sequence binaries and can be  broadly separated into
two   groups    depending   on   the   current    orbital   separation
\citep[e.g.][]{farihietal10-1}.  The first  group is  formed by  close
WDMS   binaries  that   evolved  through   a  common   envelope  phase
\citep{iben+livio93-1,  webbink07-1}.   These  systems   are  commonly
referred  to as  post-common envelope  binaries or  PCEBs and  contain
$\sim$25\,per cent  of   the   entire   WDMS   binary   population
\citep{willems+kolb04-1,  nebotetal11-1}. The  remaining $\sim$75\,per cent
 are WDMS  binaries that did not evolve through  a common envelope
phase. The orbital separation of these  systems is roughly the same as
the orbital separation of the  initial main sequence binary from which
they descend.

Thanks to the Sloan  Digital Sky Survey \citep[SDSS,][]{yorketal00-1},
the number of WDMS binaries has increased dramatically during the last
few     years    \citep[e.g.][]{silvestrietal07-1,     helleretal09-1,
  weietal13-1}.   There are  currently 2316  WDMS binaries  identified
within the  data release 8 of  SDSS \citep{rebassa-mansergasetal13-2},
of which 205 are confirmed as PCEBs \citep[e.g.][]{schreiberetal10-1}.
This  catalogue is  the largest,  most homogeneous  and cleanest  WDMS
binary sample currently available.

The SDSS WDMS binaries and PCEBs are being used to study several different
and  important   aspects  in  modern  astrophysics:   e.g.   providing
constraints    on    theories    of    common    envelope    evolution
\citep{davisetal10-1,        zorotovicetal10-1,       demarcoetal11-1,
  rebassa-mansergasetal12-2,  camachoetal14-1} and  on  the origin  of
low-mass   white  dwarfs   \citep{rebassa-mansergasetal11-1};  testing
theoretical mass-radius  relations of  both white dwarfs  and low-mass
main    sequence    stars    \citep{nebotetal09-1,    parsonsetal10-1,
  pyrzasetal12-1, parsonsetal12-1,  parsonsetal12-2}; constraining the
rotation-age-activity  relation   of  low-mass  main   sequence  stars
\citep{morganetal12-1,  rebassa-mansergasetal13-1};  constraining  the
pairing  properties  of   main  sequence  stars  \citep{ferrario12-1};
probing   the  existence   of   circumbinary   planets  around   PCEBs
\citep{zorotovic+schreiber13-1,    parsonsetal14-1,    marshetal14-1}.
However, SDSS  WDMS binaries  suffer from important  selection effects
\citep{rebassa-mansergasetal13-2}.  Namely,  the  population  of  SDSS
systems  containing cool  white  dwarfs and  early-type companions  is
clearly under-represented.

\begin{figure*}
\centering
\includegraphics[width=0.49\textwidth]{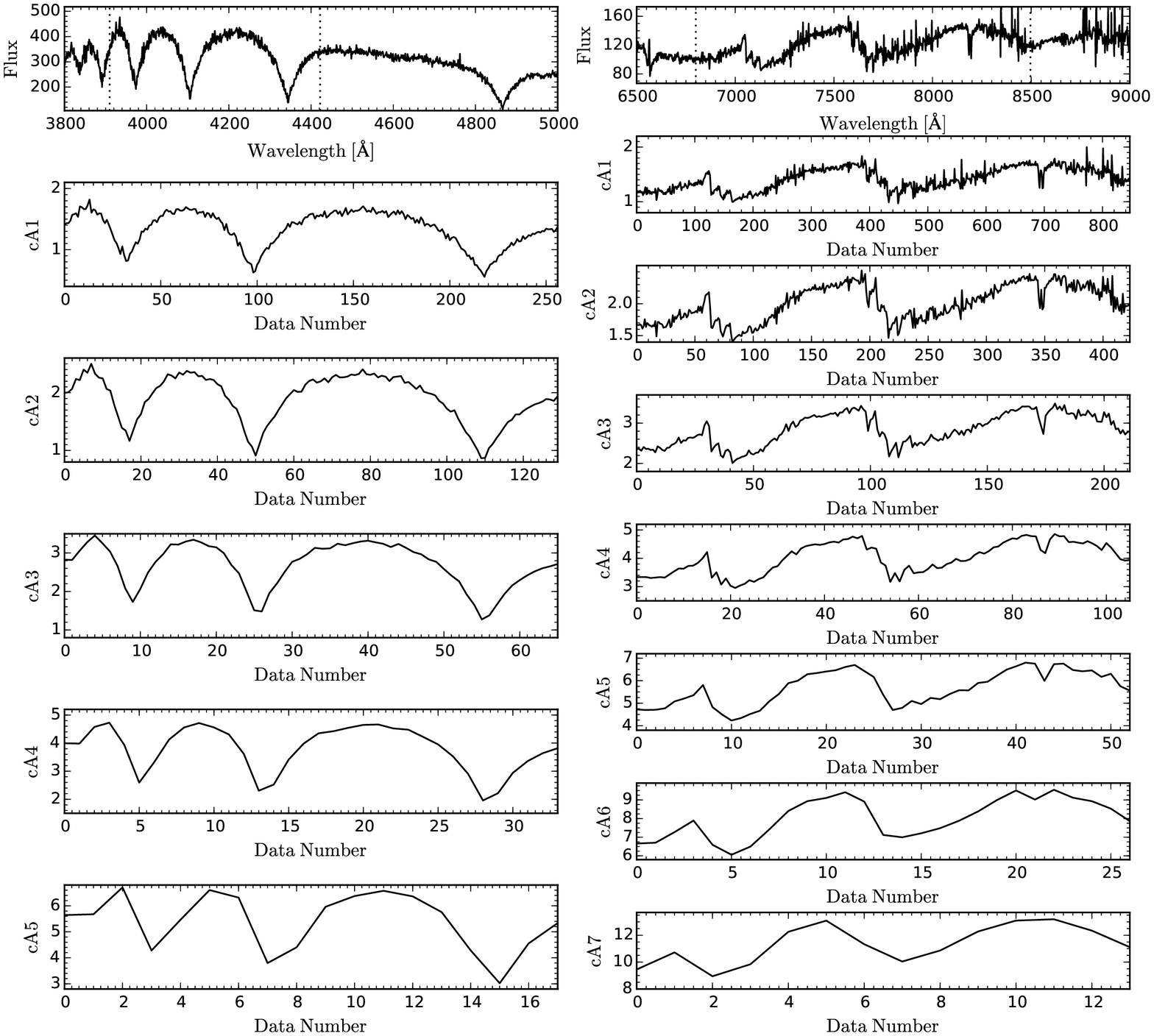}
\includegraphics[width=0.49\textwidth]{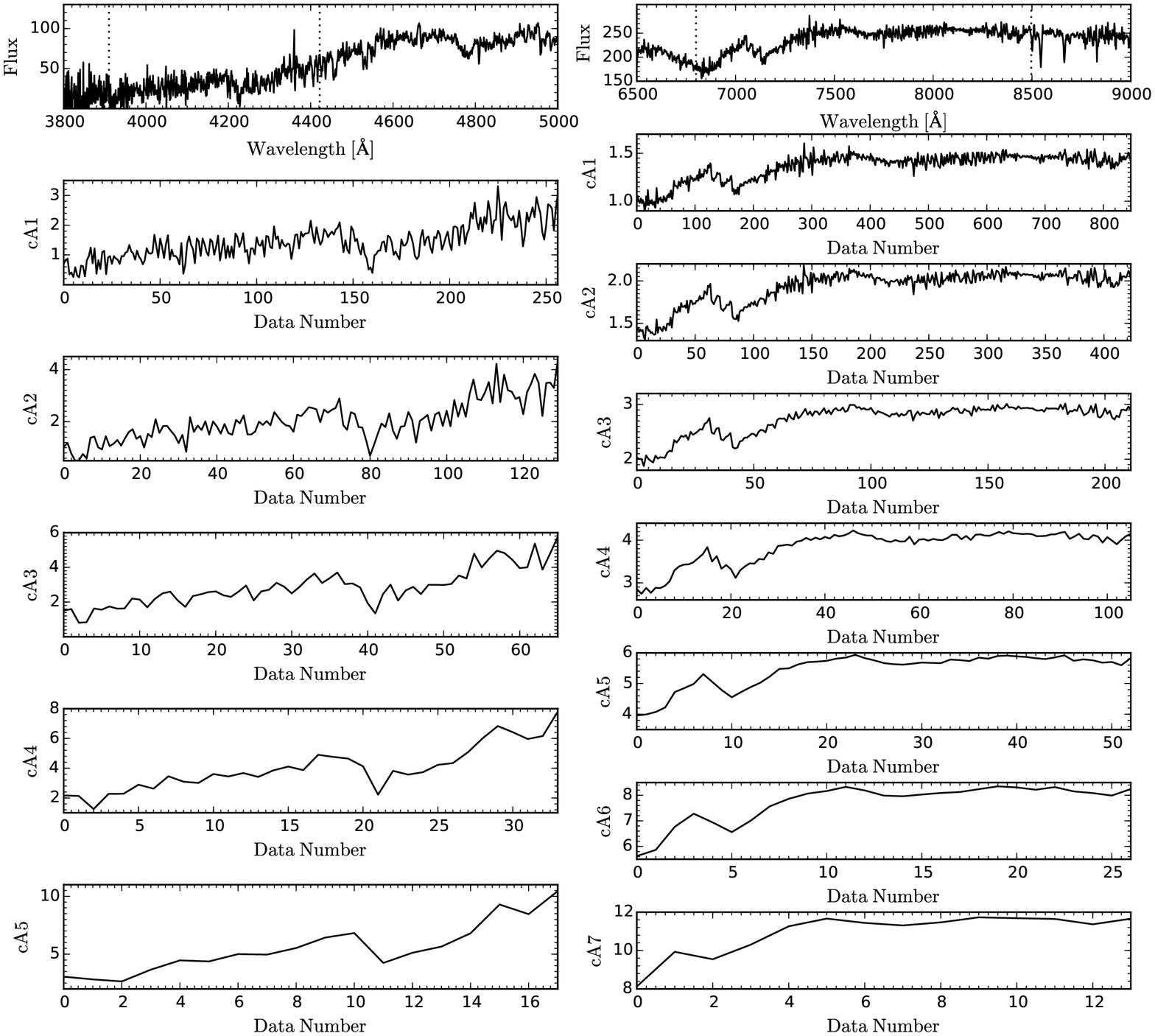}
\includegraphics[width=0.49\textwidth]{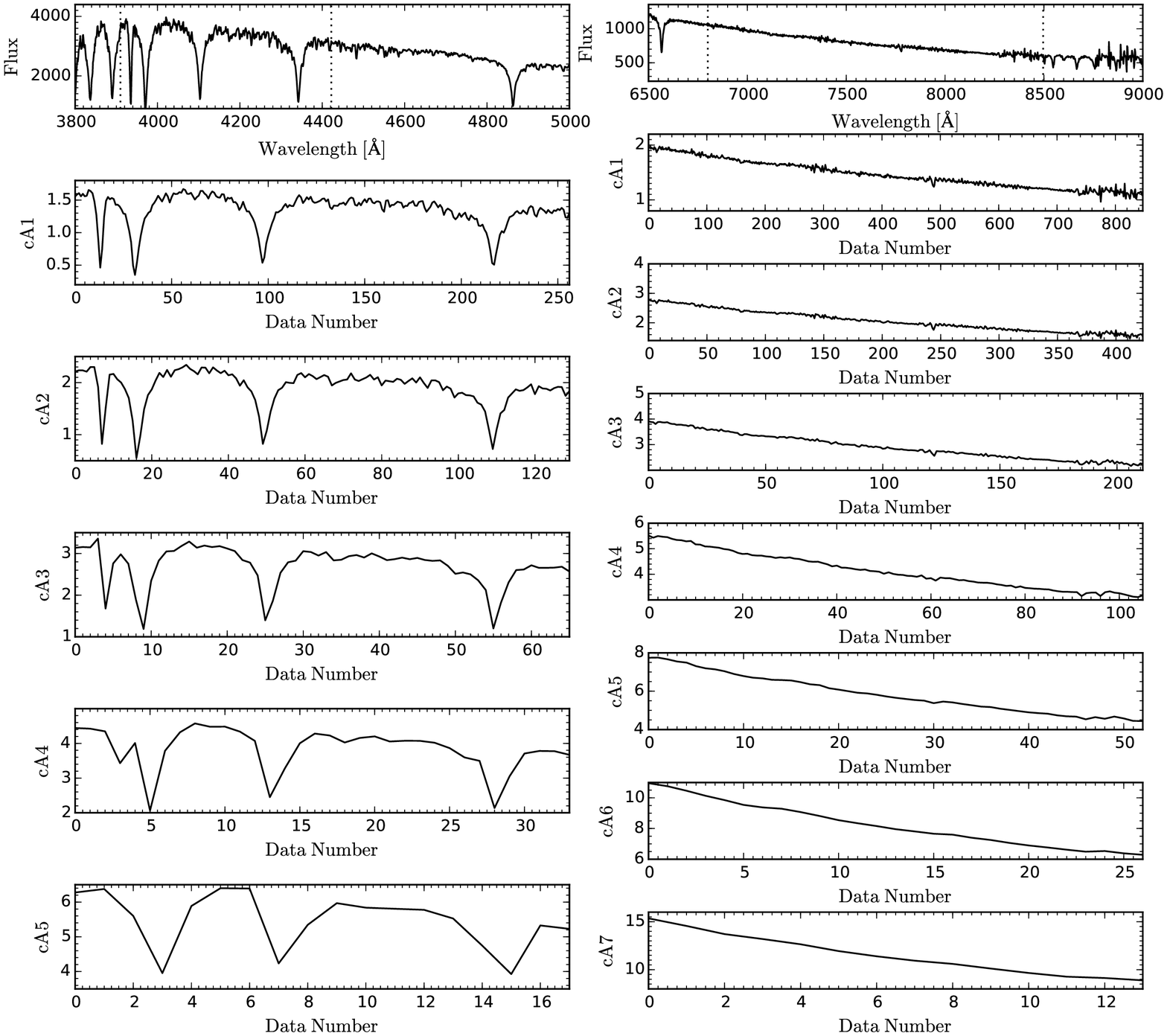}
\includegraphics[width=0.49\textwidth]{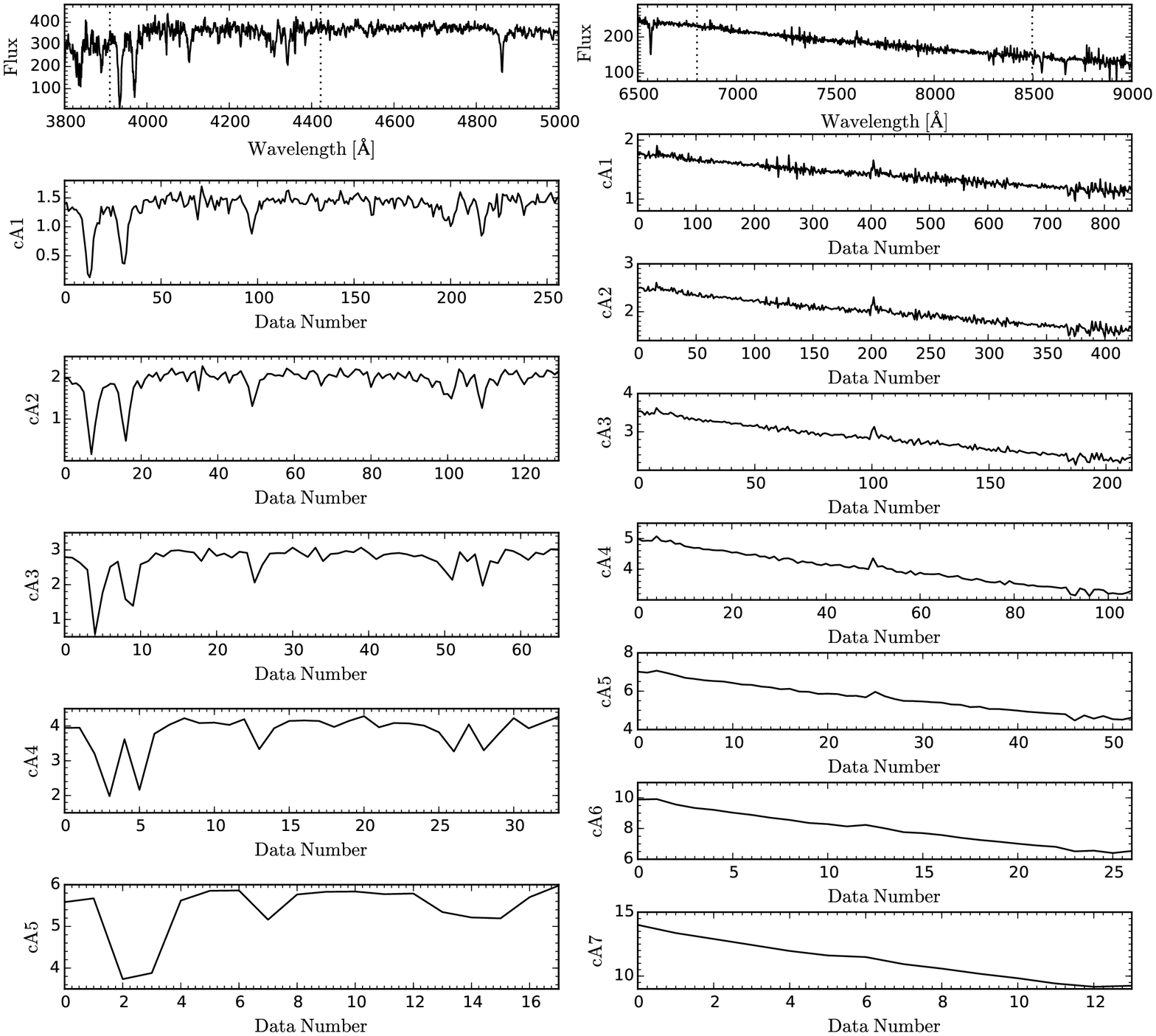}
\caption{Approximation coefficients (CA) as  a function of data number
  obtained  applying the  wavelet transform  to a  LAMOST DA/M  binary
  (J075919.43+321948.1; top left), a M star (top right panels), an A
  star (bottom  left panels), and a  G star (bottom right  panels). The
  LAMOST spectra  of each  target are  shown in  the top  panels (flux
  units   are  $10^{-17}$   erg/cm$^{-2}$\,s$^{-1}$\,\AA$^{-1}$),
  where the  vertical gray  dotted lines  define the  spectral regions
  selected to perform the wavelet transform.}
\label{fig:wavelet}
\end{figure*}

In this paper we make use of the recent, multi-faceted and large scale
survey  operated   by  LAMOST  (Large  sky   Area  Multi-Object  fiber
Spectroscopic Telescope, also named Guo Shou Jing Telescope) to extend
the search  of WDMS  binaries and  build-up the  current spectroscopic
sample.   Specifically,  we investigate  whether  or  not LAMOST  WDMS
binaries help in overcoming the selection effects incorporated by SDSS
on the current observed population.

\section{The LAMOST survey}
\label{s-lamost}

The LAMOST  is a  quasi-meridian reflecting  Schmidt telescope  located at
Xinglong   Observing  Station   in   the  Hebei   province  of   China
\citep{cuietal12-1}. It has  an effective aperture of  about 4\,meters,
and a field  of view of 5$^\circ$ in diameter. The LAMOST is exclusively
dedicated to  obtain optical  spectroscopy of celestial  objects. Each
spectral  plate refers  physically to  a focal  surface with  4000
precisely  positioned optical  fibers  to  observe spectroscopic  plus
calibration  targets  simultaneously,  equally  distributed  among  16
fiber-fed spectrographs.   Each spectrograph is equipped  with two CCD
cameras of blue and red  channels that simultaneously provide blue and
red spectra of the 4000 selected targets, respectively.

From 2012 September, LAMOST began its five-year regular survey, before
that there  was a two-year  commissioning survey and a  one-year pilot
survey.   The  LAMOST  regular  survey  consists  of  two  main  parts
\citep{zhaoetal12-1}.   The first  part is  the LAMOST  Extra-Galactic
Survey (LEGAS) of  galaxies to study the large scale  structure of the
Universe.   The second  part  is the  LAMOST  Experiment for  Galactic
Understanding  and  Exploration  (LEGUE)  Survey  of  the  Milky  Way,
developed  to  study  the  structure   and  evolution  of  the  Galaxy
\citep{dengetal12-1}. LEGUE  is subdivided  into three  surveys which
follow  independent  selection  criteria  for  follow-up  observations
\citep{carlinetal12-1, chenetal12-1,  liuetal14-1}: the  spheroid, the
disk, and the galactic anti-center.

The current  data product  of LAMOST  is data  release 1  (DR1).  This
includes 2\,204\,860 spectra, 717\,660 of  which were observed during the
pilot survey \citep{zhaoetal12-1, luoetal12-1}.  The number of spectra
classified as star, galaxy, quasar and unknown (mainly due to the poor
quality  of the  spectra) are  1\,944\,406, 12\,082,  5\,017, and  243\,355
respectively. The resolving power of  the LAMOST spectra is $\sim$1800
and covers  the $\sim$3800--9000$\lambda\lambda$ wavelength  range. 
Because of the lack of a network of photometric standard stars, the flux
calibration of LAMOST spectra  is relative \citep{songetal12-1}, which
means  that even  though  the  overall shape  of  the spectral  energy
distribution of  the targets  is well  sampled by  the flux-calibrated
spectra,  the  level  of  the   absolute  flux  may  not  be  entirely
accurate. In addition, reddening of the standard stars may also affect
the flux  calibration, especially  when the  standard stars  have very
different spatial positions  and/or reddening from other  stars in the
same  spectrograph.  A  considered LAMOST  spectrum  is  unambiguously
identified by  the MJD  (modified Julian date)  of the  observation, a
plate  identifier, a  spectrograph identifier  and a  fiber identifier
(see  \citealt{luoetal12-1}  for the  description  of  LAMOST 1D  FITS
files).

\section{The LAMOST DR1 WDMS binary sample}

In this  section we describe  in detail our procedure  for identifying
LAMOST  WDMS binaries,  estimate  the completeness  of the  identified
sample, and evaluate how efficient  the selection criteria employed by
the  LAMOST surveys  are  in targeting  WDMS  binaries for  follow-up
observations.

\subsection{Identification of LAMOST WDMS binaries}

The  first   catalogue  of  LAMOST   WDMS  binaries  is   provided  by
\citet{renetal13-1} who, based on spectrophotometric color selection,
identified  28 systems  from the  LAMOST pilot  survey.  However,  the
relative     flux     calibration     of    the     LAMOST     spectra
(Section\,\ref{s-lamost}) together  with the high  Galactic extinction
and     reddening    \citep[][especially     in    the     anti-center
  region]{liuetal14-1}  incorporate  systematic uncertainties  to  the
inferred spectrophotometric  magnitudes, which are expected  to affect
the  efficiency  of  this  method  in  selecting  WDMS  binaries.   An
additional technique  for identifying WDMS binaries  based on $\chi^2$
template fitting  was developed  by \citet{rebassa-mansergasetal10-1},
which resulted in  the largest and most homogeneous  catalogue of SDSS
WDMS  binaries  known  to  date.   However,  the  efficiency  of  this
technique in finding LAMOST WDMS binaries  may also be affected by the
un-reddened  LAMOST spectra,  as  the $\chi^2$  values would  increase
considerably.

\begin{figure*}
\centering
\includegraphics[width=0.7\textwidth,angle=-90]{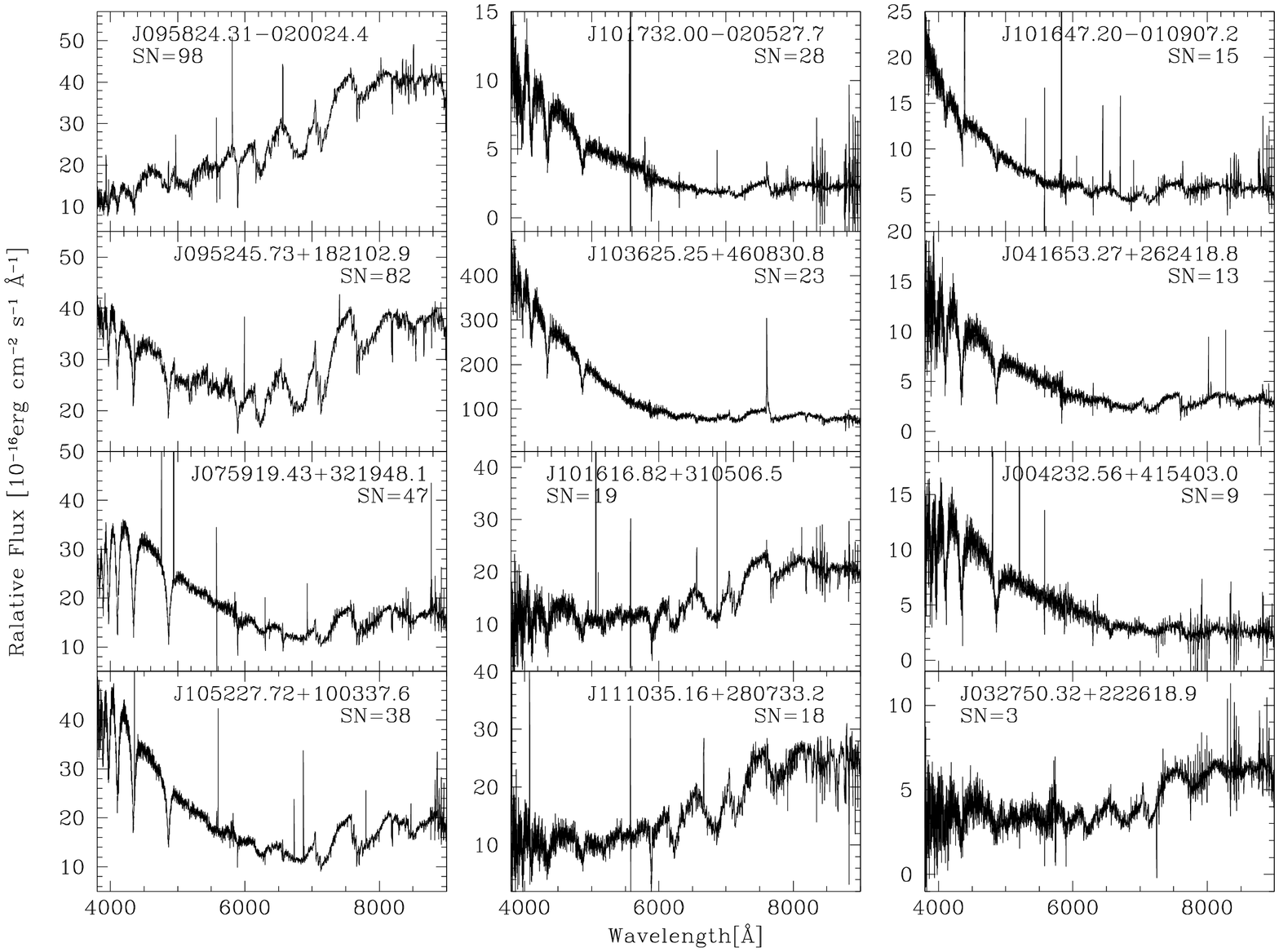}
\caption{Example  spectra  of  12  LAMOST DA/M  binaries,  ordered  in
  appearance  according to  the  S/N  of the  LAMOST
  spectra. The  complete catalogue is available at the CDS.}
\label{fig:spec}
\end{figure*}

We  here developed  a novel  routine  based on  the wavelet  transform
\citep[WT,  e.g.][]{chui92-1}  to  identify WDMS  binaries  among  the
$\sim$2  million  DR1   LAMOST  spectra.   The  WT   is  an  efficient
multi-resolution time-frequency  analysis tool, which is  very popular
in many areas of signal processing, and also has been extensively used
in        astronomical        spectral       feature        extraction
\citep[e.g.][]{starcketal97-1, Li2012}.   The analysis unit of  the WT
is  the  local flux  of  the  spectrum,  i.e.  the  selected  spectral
features.  Thus, for a given  spectrum, the WT recognises the spectral
features   rather  than   the  global   continuum,  and   consequently
uncertainties  in  the flux  calibration  and  reddening effects  are
ignored to  some extent.   The WT  is therefore  a suitable  method to
search for LAMOST WDMS binaries.

In a WT, the considered spectral  feature of a spectrum (a WDMS binary
spectrum in our case) is  decomposed into approximation signals, often
referred to  as approximation coefficients, and  the wavelength values
under the considered  spectral region are converted  into data points.
The   number   of   data   points   is   defined   mathematically   as
$(N_{2}-N_{1})/2$,  where $N_{2}$  and  $N_{1}$ are  the superior  and
inferior  sampling data  point limits  where the  WT is  applied.  The
outcome of a WT can be  hence considered (for comparative purposes) as
a smooth version  of the spectrum.  This decomposition  process can be
iterated by decomposing the approximation coefficients into successive
approximation coefficients (thus  reducing by half the  number of data
points   in  each   iteration)  until   the  decomposition   level  is
satisfactory.  This occurs when the spectral features of a WDMS binary
spectrum  can  be  identified  in the  approximation  coefficients  as
compared  to a  non-WDMS binary  spectrum in  which the  approximation
coefficients are  dominated by  continuum emission and/or  by spectral
features different from those of WDMS binaries.

Two obvious spectral features can be identified in the spectra of WDMS
binaries: the  Balmer lines in the  blue band (arising from  the white
dwarf if the  atmosphere is hydrogen-rich, i.e. a DA  white dwarf) and
the molecular  absorption bands in  the red band (typical  features of
late   type,   M   dwarf    companions).    We   thus   selected   the
3910--4422$\lambda\lambda$ range  in the blue band  (which includes the
H$\gamma$,   H$\delta$  and   H$\epsilon$   Balmer   lines)  and   the
6800--8496$\lambda\lambda$ range in the red band (which samples a large
number  of TiO  and VO  molecular bands)  as the  spectral regions  to
perform the  WT. This  obviously limited our  search to  WDMS binaries
containing a  DA white dwarf and  a M dwarf companion,  hereafter DA/M
binaries.   However,  reliable  stellar  parameters  (fundamental  for
characterising the  LAMOST population) can  only be obtained  for DA/M
binaries  \citep{rebassa-mansergasetal10-1}.  Moreover,  DA/M binaries
are    the   most    common   ($\sim$80\,per cent)    among   WDMS    binaries
\citep{rebassa-mansergasetal13-2}.

We determined the appropriate number  of WT iterations to be performed
to the two selected spectral  regions as followed.  First, we visually
inspected all  spectra classified as  white dwarf,  A star, K  star, M
star  and Double  (i.e.   binary  star) by  the  LAMOST pipelines  and
identified 82  clear DA/M binary  spectra, a list  we refer to  as the
test sample  hereafter. We note  that we  inspected these  spectral type
groups simply  because DA/M binaries are  likely typed as such  by the
LAMOST pipelines.  Then,  we randomly added a large  number of spectra
classified as Galaxy,  quasar, and star (spectral type O,  B, A, F, G,
K,  M  and  white  dwarf;  which   are  not  DA/M  binaries)  to  this
list. Finally, we  applied successive WT iterations to  all spectra in
our  test sample  and  evaluated when  the approximation  coefficients
became blurred or unrecognisable for our DA/M binaries.  We found this
happened for  the majority of cases  at the 6th iteration  in the blue
band and  the 8th iteration in  the red band.  We  therefore fixed the
number of iterations as five and seven  in the blue and the red bands,
respectively.  This  corresponds to 18  data number points  left after
the 5th iteration in  the blue band and 14 data  points left after the
7th iteration in  the red band.  In  Figure\,\ref{fig:wavelet} we show
the approximation coefficients as a  function of data number of points
obtained for  the successive  WT iterations applied  to a  DA/M binary
(top left),  a M star (top  right), an A  star (bottom left) and  a G
star (bottom right)  in our test sample.  It becomes  obvious that the
approximation  coefficients of  the A  star  and the  DA/M binary  are
similar in the  blue band, the same as  the approximation coefficients
of the  M star  and the DA/M  binary in the  red band.   Therefore, in
order  to efficiently  select  DA/M binaries  against  other types  of
celestial objects we applied the following cuts in the blue band and red band,

\begin{table*}
\caption{\label{t-cat}  Our catalogue  of  121  LAMOST DA/M  binaries.
  White  dwarf  stellar  parameters  (effective  temperature,  surface
  gravity and mass), spectral type of the companions and distances are
  given  when available,  however  only those  with a  S/N
   higher than 12 (second  column) are considered in the analysis
  of  this paper.   Spectral  types of  -1 imply  that  no values  are
  available.    Modified   Julian   date   (MJD),   plate   (plateid),
  spectrograph (spid) and fiber (fiberid)  identifiers for each of the
  LAMOST spectra are also provided.  For completeness, we also include
  181 systems that are not considered  by us as DA/M binaries but that
  show blue and red components in their spectra.  These are flagged as
  1  in the  last column.   The  complete table  can be  found in  the
  electronic version of the paper.}  \setlength{\tabcolsep}{0.6ex}
\begin{small}
\begin{tabular}{cccccccccccccccc}
\hline
\hline
Jname  &    S/N &   MJD &       plateid       &      spid   &  fiberid   &  T$_\mathrm{eff} ^\mathrm{wd}$ &  err & log(g)$_\mathrm{wd}$ & err & M$_\mathrm{wd}$ &   err &   d &       err     &       Sp      & type    \\
       &        &       &                     &             &            &  [K]                        &      &                    &     & [M$_\odot$]     &       &  [pc]           &               &               &       \\  
\hline
J002237.90+334322.1  &   5.48   & 56255 &  VB007N33V1             &   03  &    085  &   0     &  0    &   0     &  0    &  0    &   0    &  0     &  0    &   -1  &   0 \\
J002952.22+434205.8  &   37.36  & 56266 &  M31007N41B2            &   11  &    196  &   34926 &  1011 &   8.29  &  0.19 &  0.83 &   0.11 &  245   &  34   &   2   &   0 \\
J004232.56+415403.0  &   8.75   & 55886 &  M31\_004437N404045\_M2 &   15  &    202  &   15071 &  702  &   7.08  &  0.16 &  0.26 &   0.04 &  238   &  27   &   6   &   0 \\
J004432.63+402635.2  &   11.25  & 55886 &  M31\_004437N404045\_M2 &   04  &    127  &   0     &  0    &   0     &  0    &  0    &   0    &  0     &  0    &   3   &   0 \\
J004545.96+415029.9  &   4.57   & 55863 &  M31\_011N40\_B1        &   15  &    128  &   0     &  0    &   0     &  0    &  0    &   0    &  0     &  0    &   2   &   0 \\
J005221.55+440335.4  &   15.87  & 56207 &  M31011N44B1            &   08  &    149  &   22292 &  3046 &   8.23  &  0.48 &  0.77 &   0.29 &  225   &  74   &   2   &   0 \\
J005601.44+400718.4  &   28.14  & 55886 &  M31\_004437N404045\_M2 &   06  &    151  &   0     &  0    &   0     &  0    &  0    &   0    &  0     &  0    &   2   &   0 \\
J005636.91+383332.6  &   16.69  & 55914 &  M31\_01h38\_F1         &   03  &    098  &   0     &  0    &   0     &  0    &  0    &   0    &  0     &  0    &   4   &   0 \\
J005853.40+064849.3  &   59.70  & 56240 &  EG010249N073002F       &   03  &    027  &   11565 &  247  &   7.55  &  0.25 &  0.38 &   0.11 &  229   &  35   &   1   &   0 \\
J011447.51+254229.6  &   16.66  & 56267 &  EG011927N271550M01     &   02  &    241  &   35740 &  628  &   7.81  &  0.12 &  0.57 &   0.06 &  140   &  12   &   1   &   0 \\
\hline
\end{tabular}
\end{small}
\end{table*}

\begin{table*}
\centering
\caption{\label{t-mag} SDSS  or Xuyi  magnitudes (when  available) and
  coordinates of our 121 LAMOST DA/M binaries.  The complete table can
  be   found    in   the    electronic   version   of    the   paper.}
\setlength{\tabcolsep}{1ex}
\begin{small}
\begin{tabular}{cccccccccc}
\hline
\hline
Jname  & ra[deg] & dec[deg] & $u$ & $g$ & $r$ & $i$ & $z$ & SDSS/Xuyi \\
\hline
J002237.90+334322.1  &      5.65794  &     33.72280  &     19.56  &   17.35  &    15.98 &     14.58 &     13.87 &      SDSS \\
J002952.22+434205.8  &      7.46761  &     43.70163  &     0      &   0      &    0     &     0     &     0     &      0    \\
J004232.56+415403.0  &     10.63567  &     41.90083  &     17.35  &   17.18  &    17.43 &     17.40 &     17.09 &      SDSS \\
J004432.63+402635.2  &     11.13595  &     40.44311  &     19.80  &   19.03  &    17.96 &     16.77 &     16.08 &      SDSS \\
J004545.96+415029.9  &     11.44152  &     41.84165  &     0      &   0      &    0     &     0     &     0     &      0    \\
J005221.55+440335.4  &     13.08981  &     44.05985  &     17.05  &   16.46  &    15.62 &     14.61 &     13.95 &      SDSS \\
J005601.44+400718.4  &     14.00604  &     40.12179  &     18.81  &   18.13  &    17.23 &     16.35 &     15.88 &      SDSS \\
J005636.91+383332.6  &     14.15381  &     38.55906  &     0      &   0      &    0     &     0     &     0     &      0    \\
J005853.40+064849.3  &     14.72252  &      6.81371  &     17.30  &   17.18  &    17.46 &     17.64 &     17.73 &      SDSS \\
J011447.51+254229.6  &     18.69799  &     25.70823  &     16.96  &   17.18  &    17.49 &     17.31 &     16.85 &      SDSS \\
\hline
\end{tabular}
\end{small}
\end{table*}

\begin{equation}
\begin{array}{l}
f_b[i_1] < max(f_b[i_1 + 1], f_b[i_1 + 2])\ \| \\
(f_{b}[i_1] < f_{b}[i_1 + 2]\ \&\ f_b[i_1] < max(f_{b}[i_1 - 1], f_{b}[i_1 - 2]))\ \& \\
(f_{b}[i_2] < f_{b}[i_2 - 2]\ \&\ f_{b}[i_2] < max(f_{b}[i_2 + 2], f_{b}[i_2 + 3]))\ \& \\
(f_{b}[i_3] < f_{b}[i_3 - 3]\ \&\ f_{b}[i_3] < f_{b}[i_3 + 1])
\end{array}
\end{equation}

\noindent and

\begin{equation}
\begin{array}{l}
f_{r}[j_1] > f_{r}[j_1 + 1]\ \& (f_{r}[j_2] > f_{r}[j_2 - 3]\ \&\ \\
f_{r}[j_2] > f_{r}[j_2 + 2])\ \&\, min(f_{r}[j_3], f_{r}[j_3 - 1]) < \\
max(f_{r}[j_3 - 2], f_{r}[j_3 - 3], f_{r}[j_3 - 4])
\end{array}
\end{equation}

\noindent respectively,  where  $f_b$  and   $f_{r}$  are  the
approximation coefficients in the blue  and the red bands respectively
at the corresponding data  number $i$ and $j$, and $i_1$  = 3, $i_2$ =
7, $i_3$ = 15, and $j_1$ = 1, $j_2$  = 5, $j_3$ = 13. In the blue band
these  equations select  the  DA/M  binary and  the  A  star shown  in
Figure\,\ref{fig:wavelet}  and exclude  the  G star  and  the M  star,
whilst in the red band they select  the DA/M binary and the M star but
exclude  the A  star  and the  G  star.  Thus,  only  the DA/M  binary
survives  the applied  cuts  in  Figure\,\ref{fig:wavelet}.  From  the
complete test  sample, these  cuts efficiently selected  79 of  the 82
DA/M binaries . However, we note that we applied  a large number of different cuts,
and only  those provided  here resulted in  the largest  number of
DA/M binaries identified from our test sample.

Applying  the WT  and above  define cuts  to all  $\sim$2 million  DR1
LAMOST spectra resulted in a sample of 8543 selected spectra, which we
visually inspected. This resulted in the identification of 130 genuine
DA/M binary spectra of 118  unique systems (nine objects were observed
twice by  LAMOST, and one was  observed four times).  To  this list we
added the three  DA/M binary spectra missed from the  test sample (one
is a  different spectrum  of a DA/M  binary already  identified), thus
increasing the total number of LAMOST  DA/M binaries to 120 (133 total
spectra). A  cross-correlation of our  list of DA/M binaries  with the
catalogue  by   \citet{rebassa-mansergasetal13-2}  revealed   that  41
systems  were  previously observed  by  SDSS,  which implies  we  have
identified 79  new systems in this  work. We also find  a large number
(180) of  systems displaying both  a red and  a blue component  in the
LAMOST spectra,  but that  we do  not consider  as DA/M  binaries. The
spectra of these objects are either too noisy, or are likely to be the
result of the superposition of two stars  in the line of sight, or the
result  of  white  dwarfs  (M-dwarfs) located  close  to  very  bright
M-dwarfs (white  dwarfs or A-stars)  causing scattered light  to enter
the spectroscopic fiber, i.e.  an apparent two-component spectrum.

Finally, we  compared our list  of 120 systems  with the sample  of 28
LAMOST   DA/M    binaries   from   the   pilot    survey   of   LAMOST
\citep{renetal13-1}.  Our WT method finds 26 of the 28 listed targets.
However,  four  of  these  (J052529.17+283705.4,  J052531.26+283807.6,
J052531.33+284549.4, and J120024.56+292310.3) are  not included in our
sample as we do not consider them DA/M binaries. Of the two objects we
missed, one (J132417.76+280755.8) is a DA/M and we add it to our list,
the other  (J150626.53+275925.2) a  DA/M binary candidate.   The total
number of LAMOST DA/M binaries thus increases to 121 (the total number
of LAMOST  spectra is 134) and  the final number of  new DA/M binaries
found in  this work,  i.e.  systems  that were  not observed  by SDSS,
increases  to  80.   Table\,\ref{t-cat}  lists  all  our  LAMOST  DA/M
binaries.   For completeness,  we also  include in  this list  the 181
systems displaying two  components in the LAMOST spectra  but that are
not  considered as  DA/M binaries  by us.   Table\,\ref{t-mag}
  lists the SDSS  $ugriz$ or Xuyi $gri$ magnitudes  and coordinates of
  our  121  DA/M   binaries  (the  Xuyi  survey   is  the  photometric
  counterpart of  the LAMOST survey  of the Galactic  anti-center; see
  Section\,\ref{s-lamost}).   In Figure\,\ref{fig:spec}  we show  the
LAMOST  spectra of  12  DA/M binaries  in  our catalogue  (the
  complete catalogue is available at the CDS).

\begin{figure*}
\centering
\includegraphics[width=0.49\textwidth,angle=0]{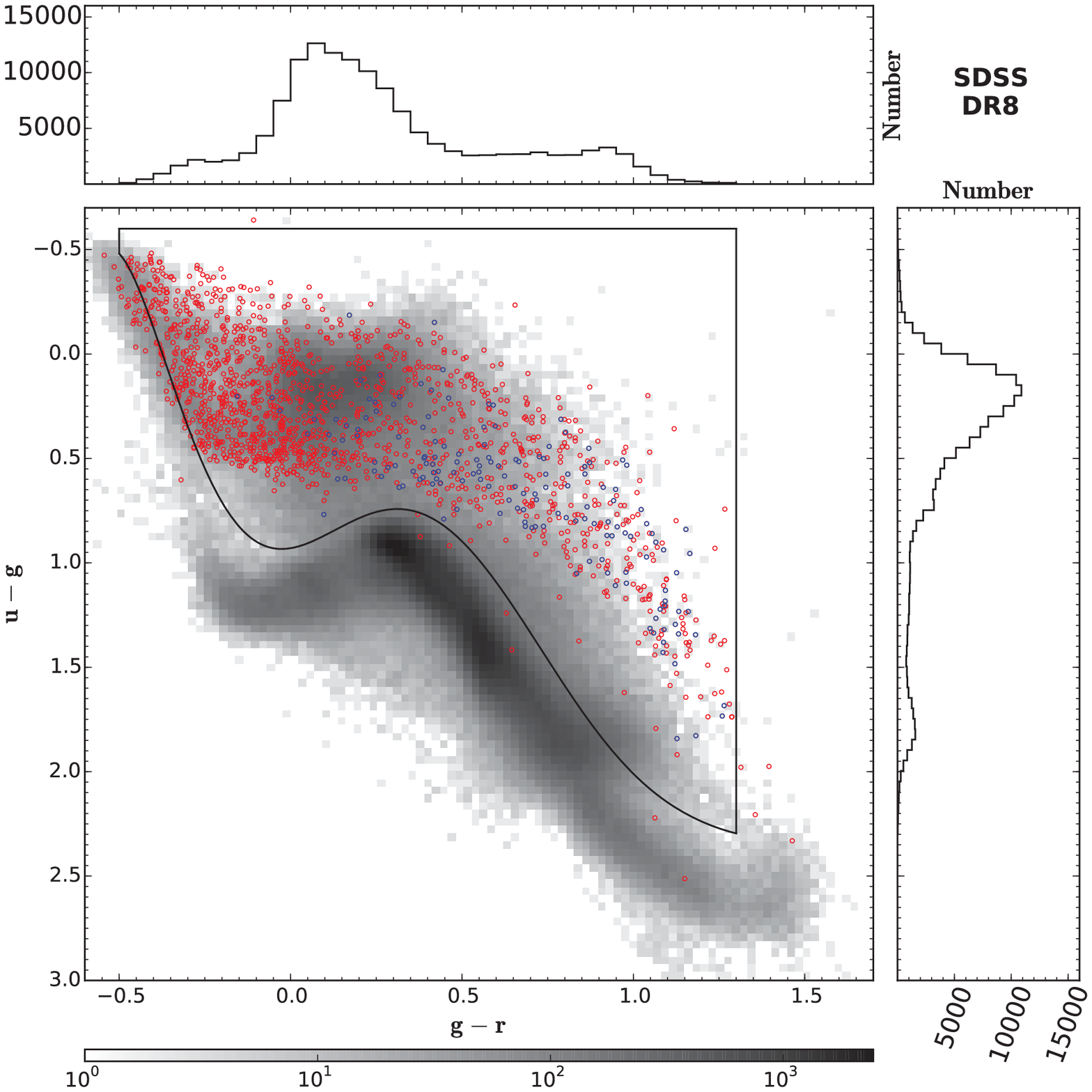}
\includegraphics[width=0.49\textwidth,angle=0]{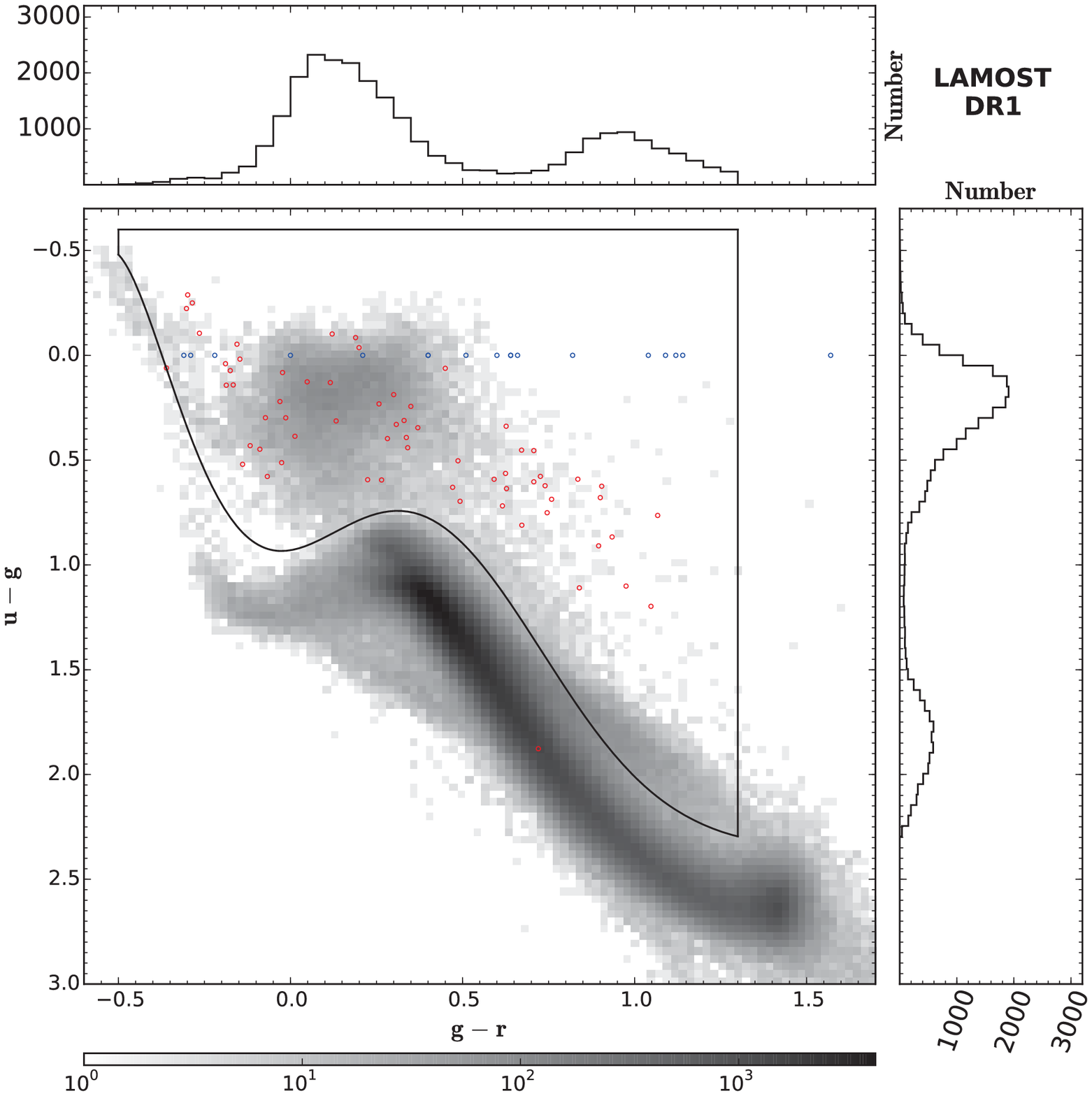}
\caption{SDSS (left) and  LAMOST (right) density maps in  the $u-g$ vs.
  $g-r$ plane. Identified DA/M binaries  are shown as red open circles
  (blue open circles represent SEGUE and GAC DA/M binaries in the left
  and  right  panels,  respectively).    The  solid  black  lines  are
  color-cuts defined for selecting DA/M binaries and excluding single
  main  sequence  stars.  The  number  distribution  of objects  as  a
  function of  $u-g$ and $g-r$ within  these cuts are provided  in the
  upper and right panels respectively.}
\label{fig:color}
\end{figure*}

\subsection{The efficiency of the wavelet transform method 
and completeness of the LAMOST WDMS binary sample}

In order to  evaluate the efficiency of the WT  method in finding DA/M
binaries as  well as to estimate  the completeness of the  LAMOST DA/M
binary  sample, i.e.   the  fraction of  DA/M  binaries we  identified
respect to the total number observed,  we applied the WT method to all
$\sim$1.8 million SDSS DR\,8 spectra. Our purpose here is to check how
many     of     the     1819      SDSS     DR\,8     DA/M     binaries
\citep{rebassa-mansergasetal13-2} can be identified by the WT routine.
After applying  the WT to  all SDSS  DR\,8 spectra we  obtained 27\,060
WDMS  binary  candidate spectra,  and  after  a visual  inspection  we
identified a total  of 1646 systems as DA/M  binaries.  Therefore, the
WT recovered $\sim$90\,per cent of  the SDSS DA/M binary sample.  Given
that  the SDSS  WDMS binary  catalogue  is $\ga$98\,per cent  complete
\citep{rebassa-mansergasetal13-2},   the   exercise   performed   here
strongly suggests the  WT should select also $\sim$90\,per cent of all
DA/M binaries  observed by LAMOST.   The spectra  of the 173  SDSS DR8
DA/M binaries the WT failed to  identify miss important data points in
the areas in which the WT is  applied, or are clearly dominated by one
of  the  two stellar  components,  or  are  spectra of  extremely  low
signal-to-noise  ratio (S/N)  for a  clear  identification  of the  spectral
features of the two stars.  We  expect the $\sim$10\,per cent of LAMOST
DA/M  binary   spectra  missed   by  the   WT  to   be  of   the  same
characteristics.

We conclude  that the WT is  an efficient method for  identifying DA/M
binaries and that the completeness of the LAMOST DA/M binary sample is
$\sim$90\,per cent.

\subsection{The efficiency of the LAMOST target selection algorithm 
in selecting WDMS binaries}
\label{s-selection}

The number of  spectra released by SDSS DR\,8  ($\sim$1.8 million) and
LAMOST DR\,1 ($\sim$2 million) are similar.  However, the total number
of  DA/M binaries  in LAMOST  DR\,1 (121)  is more  than one  order of
magnitude lower  than in SDSS  DR\,8 (1819).  This difference  is very
likely  a  consequence  of  the different  target  selection  criteria
employed by the two surveys. We investigate this hypothesis here.
 
In Figure\,\ref{fig:color} (left  panel) we show the  $u-g$ vs.  $g-r$
density map  for 497\,354  SDSS DR\,8  sources with  available spectra,
clean photometry, and photometric errors of less than 0.05 magnitudes.
In the  right panel of the  same Figure we represent  the same density
map for 587\,988  DR\,1 LAMOST targets with available  SDSS DR\,8 $ugr$
magnitudes, where again  we select only objects  with clean photometry
and photometric  errors below  0.05.  We  used this  particular color
diagram  because single  main sequence  stars can  be
easily excluded in  the $u-g$ vs. $g-r$ plane, which  allows us to analyse
the impact  of the  selection criteria in  targeting DA/M  binaries in
both quasar dominated and quasar free color areas by the two surveys.
The  color  cut   defined  by  \citet{rebassa-mansergasetal13-2}  for
excluding single  main sequence  stars is  illustrated as  black solid
lines in  both panels  of Figure\,\ref{fig:color}, together  with 1523
SDSS DR\,8  and 70 LAMOST DR\,1  DA/M binaries that satisfy  the above
photometric conditions (red open  circles).  For comparative purposes,
SEGUE (the SDSS Extension  for Galactic Understanding and Exploration,
\citealt{yannyetal09-1}) DA/M  binaries, which  contain systematically
cooler     white    dwarfs     and     earlier    type     secondaries
\citep{rebassa-mansergasetal12-1}, are displayed  as open blue circles
in the left panel of  Figure\,\ref{fig:color}. It is worth noting that
the LAMOST  survey of the  Galactic anti-center, i.e.  the  GAC survey
(Section\,\ref{s-lamost}), does not overlap with SDSS DR\,8, therefore
the  vast   majority  of   targets  observed   are  not   included  in
Figure\,\ref{fig:color} because of lack of $ugr$ magnitudes.  Thus, we use
the  $g-r$  colors obtained  from  Xuyi  photometry (the  photometric
counterpart of  the GAC  survey, \citealt{liuetal14-1}) of  all LAMOST
DA/M  binaries that  were  observed  as part  of  the  GAC survey  and
illustrate  them  as   blue  open  circles  in  the   right  panel  of
Figure\,\ref{fig:color}, where  we assume a  $u-g$ color of 0  because of
the lack of Xuyi $u$-band photometry.

Inspection of  Figure\,\ref{fig:color} reveals  that outside  the DA/M
binary color space  area the LAMOST and SDSS density  maps are rather
similar, the only difference being that LAMOST observed a considerably
larger number of  main sequence stars.  Within the DA/M  binary box it
becomes  clear  that SDSS  observed  approximately  half an  order  of
magnitude more sources  than LAMOST.  This can clearly be  seen in the
number distribution of targets as a function of $u-g$ (top panels) and
$g-r$ (right panels),  and has an easy explanation: whilst  one of the
main  drivers of  SDSS has  been targeting  galaxies and  quasars for
spectroscopy         \citep{straussetal02-1,         richardsetal02-1,
  adelman-mccarthyetal08-1} (see the prominent peaks at $g-r\simeq0.2$
and $u-g\simeq0.4$),  LAMOST is a dedicated  Galactic survey targeting
mainly           stars           \citep[][          see           also
  Section\,\ref{s-lamost}]{zhaoetal12-1}.    It   is   therefore   not
surprising  that the  number  of  DA/M binaries  observed  by SDSS  is
considerably larger than the number observed by LAMOST.

Independently of the total number of DA/M binaries observed by the two
surveys, a different  and more important issue in the  context of this
paper  is whether  the  LAMOST DA/M  binary  population has  different
properties than the one observed by SDSS. Specifically, whether or not
LAMOST DA/M  binaries contain  systematically cooler white  dwarfs and
earlier-type companions,  a population  under-represented in  the SDSS
sample.    Although   we   will   investigate  this   in   detail   in
Section\,\ref{s-comparison}, it is worth noting that $\sim$70\,per cent
of SDSS and $\sim$60\,per cent of LAMOST DA/M binaries are concentrated
in quasar  dominated color  areas ($g-r  \la 0.7$)  representative of
systems containing  hot white dwarfs  (open circles in both  panels of
Figure\,\ref{fig:color}).  As expected, SEGUE DA/M binaries (blue open
circles;   left   panel    of   Figure\,\ref{fig:color})   which   are
intrinsically  dominated  by  cooler  white  dwarfs  and  earlier-type
companions, generally populate  color areas that do  not overlap with
those of quasars ($g-r \ga 0.7$). Interestingly, this is also the case
for $\sim$50\,per cent  of LAMOST  DA/M binaries  observed by  the GAC
survey (blue  open circles  in Figure\,\ref{fig:color},  right panel).
Therefore the subpopulation  of LAMOST GAC DA/M  binaries may deserve
further attention.

We conclude that the low number of DA/M binaries observed by LAMOST as
compared to  SDSS is a  simple consequence  of the differences  in the
target  selection  criteria of  both  surveys.   We also  find  strong
indications for a large population  of LAMOST DA/M binaries containing
hot white dwarfs,  which indicates the LAMOST surveys may  not help in
overcoming the selection effects incorporated  by SDSS. We will discuss
this in detail in Section\,\ref{s-comparison}.

\section{Stellar parameters and distances}
\label{s-param}

\begin{figure*}
\includegraphics[angle=-90,width=0.49\textwidth]{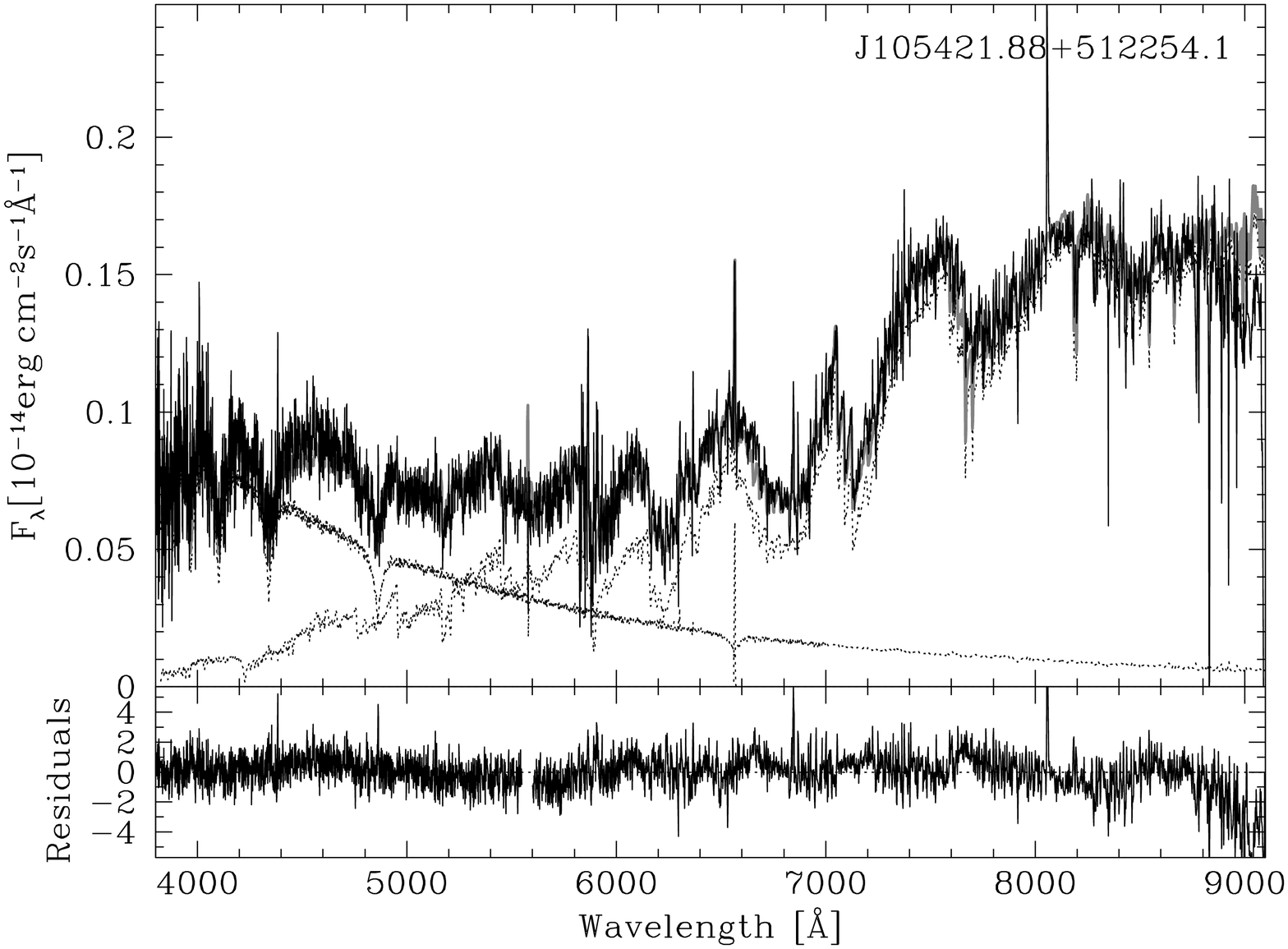}
\hfill
\includegraphics[angle=-90,width=0.49\textwidth]{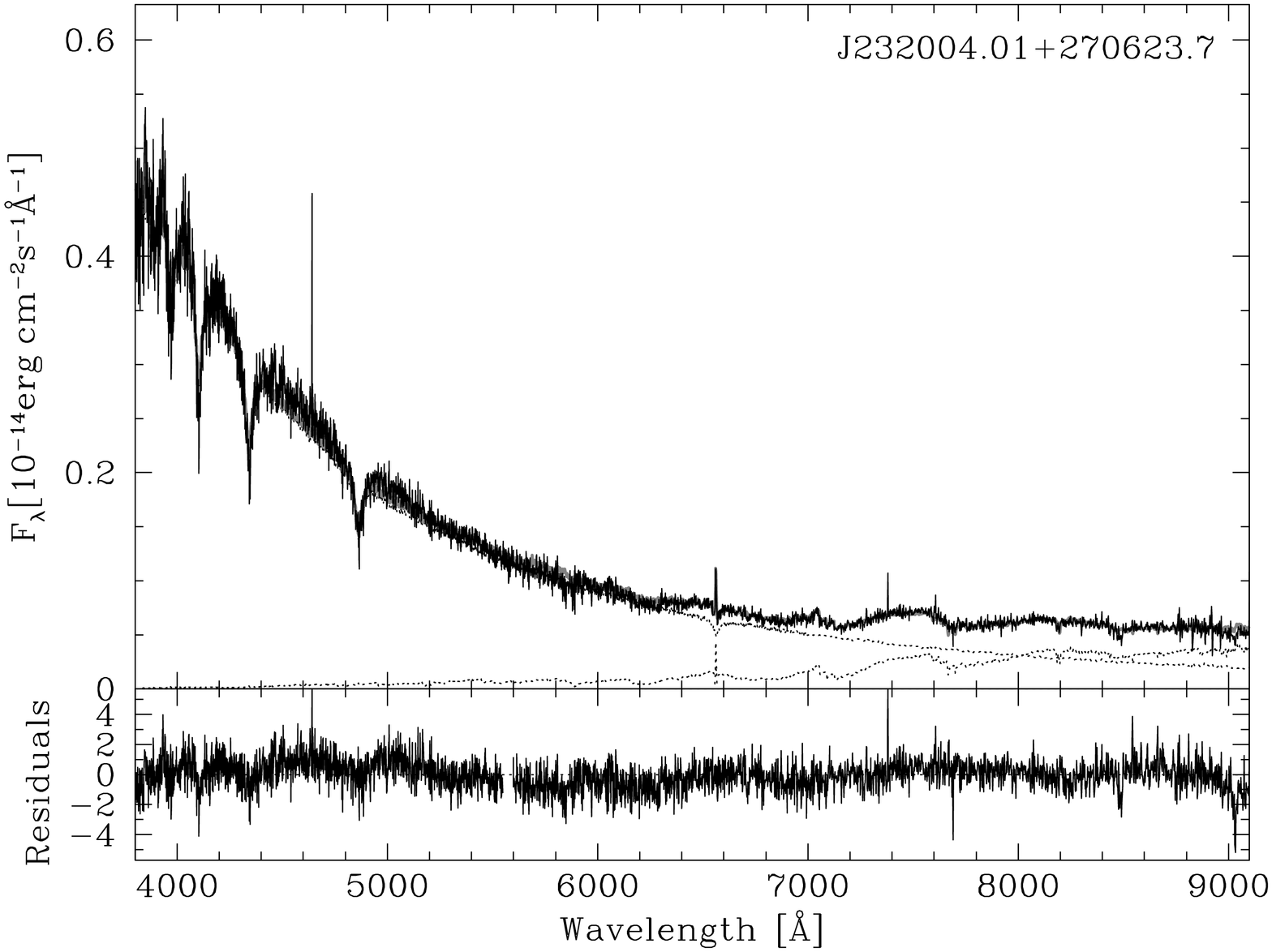}
\caption{\label{f-fullfit}  Two-component  fits  to  the  LAMOST  DA/M binary
  spectra.  Shown are examples for  objects with either the M-dwarf or
  the white dwarf dominating the LAMOST spectrum.  The top panels show
  the DA/M binary spectrum as black line,  and the two templates, white dwarf
  and M-dwarf, as  dotted lines. The bottom panels  show the residuals
  from the fit.}
\end{figure*}

\begin{figure*}
\includegraphics[width=0.49\textwidth]{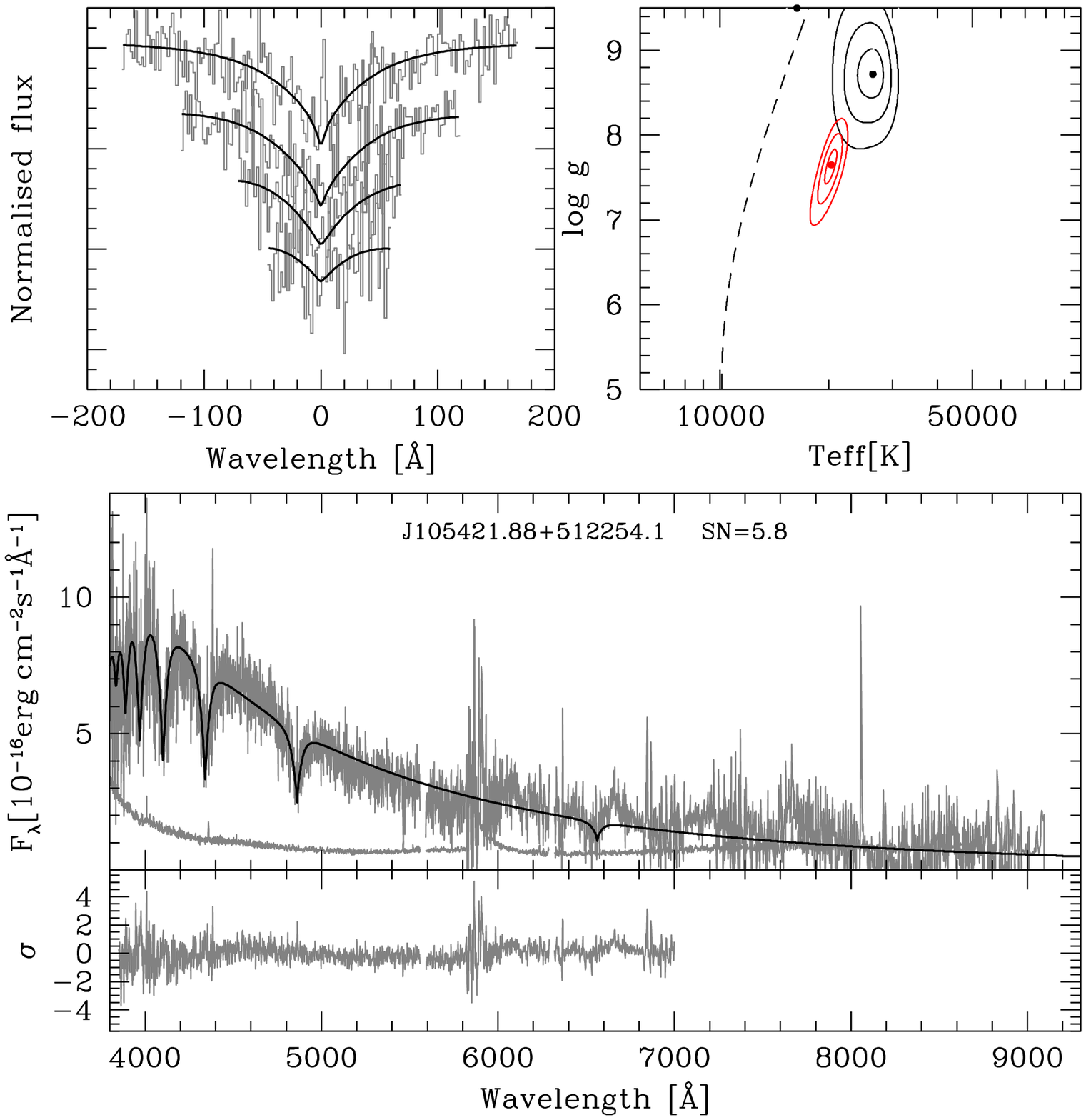}
\hfill
\includegraphics[width=0.49\textwidth]{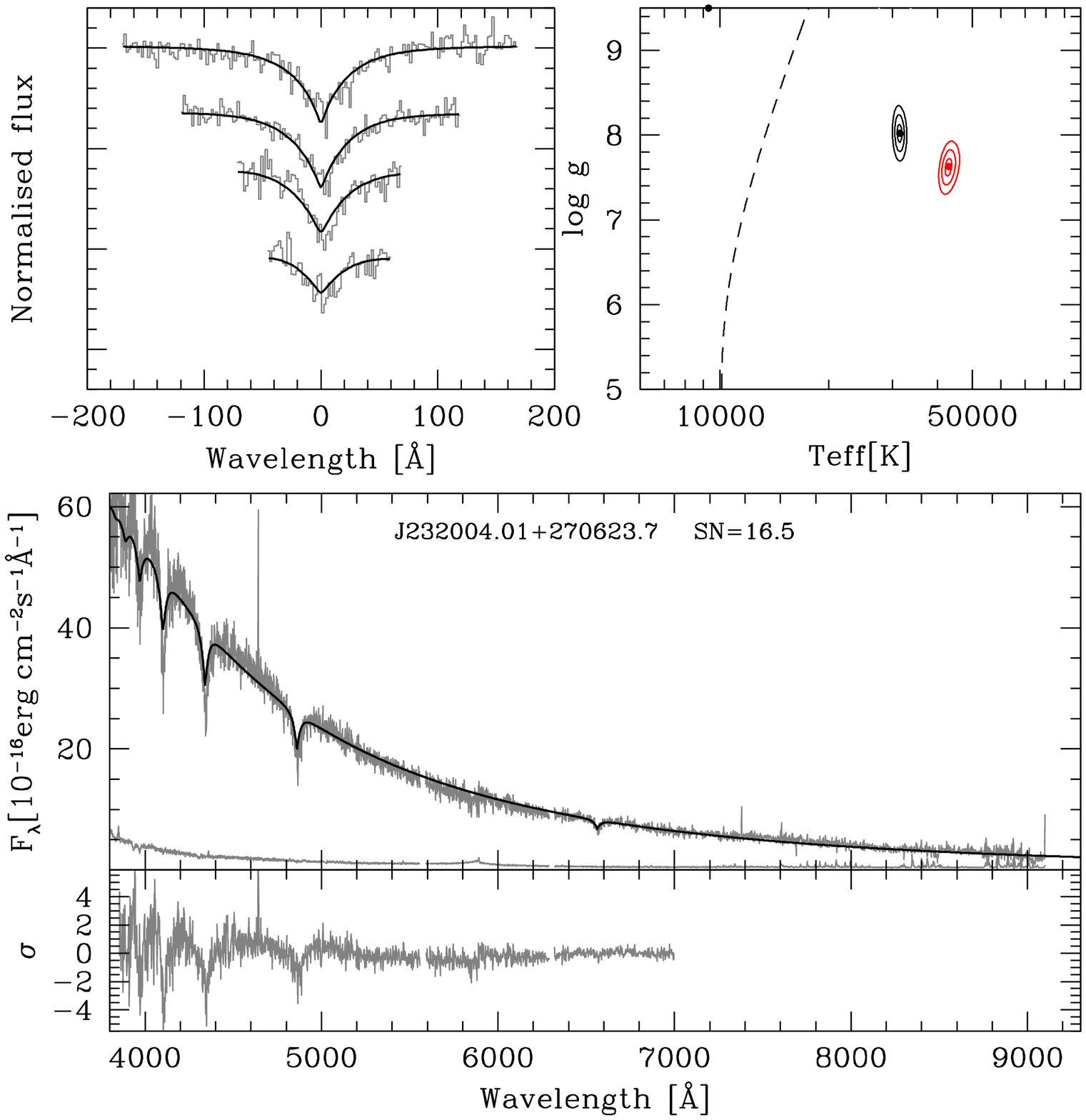}
\caption{\label{f-wdfit}  Spectral  model  fits  to  the  white  dwarf
  components    of     the    two     DA/M    binaries     shown    in
  Figure\,\ref{f-fullfit},  obtained  after subtracting  the  best-fit
  M-dwarf template.   Top left panels:  best-fit (black lines)  to the
  normalised H$\beta$ to H$\epsilon$ (gray  lines, top to bottom) line
  profiles. Top  right panels: 3,  5, and 10$\sigma$  $\chi^2$ contour
  plots in the $\Teff-\log g$ plane.   The black contours refer to the
  best line  profile fit,  the red  contours to the  fit of  the whole
  spectrum.   The  dashed line  indicates  the  occurrence of  maximum
  H$\beta$  equivalent  width.  The  best  ``hot''  and ``cold''  line
  profile solutions are  indicated by black dots, the best  fit to the
  whole  spectrum  is indicated  by  a  red  dot. Bottom  panels:  the
  residual   white  dwarf   spectra   resulting   from  the   spectral
  decomposition  and their  flux errors  (gray lines)  along with  the
  best-fit    white    dwarf    model     (black    line)    to    the
  3800--7000\,$\lambda\lambda$   wavelength   range  (top)   and   the
  residuals of the fit (gray line,  bottom).  The $\Teff$ and $\log g$
  values  listed in  Table\,\ref{t-cat} are  determined from  the best
  line profile  fit. The  fit to  the whole spectrum  is only  used to
  select between the ``hot'' and ``cold'' line fit.}
\end{figure*}

Reddening is likely  to affect the flux calibration  of LAMOST spectra
(Section\,\ref{s-lamost}), which  may affect the determination  of the
stellar  parameters of  our  DA/M binaries  (especially  if these  are
located in the Galactic anti-center  region).  To solve this potential
problem we re-flux calibrated the spectra of our 121 targets following
the routine developed by \citet{xiangetal14-1}.  This method makes use
of de-reddened flux  standard stars observed in  the same spectrograph
as the considered  targets to provide de-reddened  and flux calibrated
target   spectra.    However,   the  flux   calibration   applied   by
the \citet{xiangetal14-1} routine  is also  relative, i.e.   the overall
shape  of the  spectral energy  distribution  is well  sampled by  the
flux-calibrated  spectra  but  the  absolute flux  level  may  not  be
entirely accurate.

We      applied     the      decomposition/fitting     routine      of
\citet{rebassa-mansergasetal07-1}  to  the re-calibrated  spectra  for
measuring the stellar  parameters and estimating the  distances to our
DA/M binaries.  This method is briefly described as followed. First, a
given LAMOST DA/M binary spectrum is fitted with a two-component model
using a set  of observed M dwarf and white  dwarf templates.  From the
converged fit to each spectrum (see Figure\,\ref{f-fullfit}) we record
the  spectral  type of  the  secondary  star.   The best-fit  M  dwarf
template  is then  subtracted  and  we fit  the  residual white  dwarf
spectrum with a  model grid of DA white  dwarfs \citep{koester10-1} to
obtain its effective temperature,  $\Teff$, and surface gravity, $\log
g$ (see  Figure\,\ref{f-wdfit}).  The equivalent widths  of the Balmer
lines go  through a  maximum near  $\Teff=13\,000$\,K, with  the exact
value being a  function of $\log g$.  Therefore, $\Teff$  and $\log g$
determined from Balmer line profile  fits are subject to an ambiguity,
often referred  to as  ``hot'' and ``cold''  solutions, i.e.   fits of
similar quality can  be achieved on either side of  the temperature at
which the maximum  equivalent width is occurring.  We  use the $\Teff$
and $\log g$  obtained from the fits to the  whole spectrum, continuum
plus lines (excluding the region  above 7000\,\AA, where residual
contamination  from  the secondary  star  subtraction  is largest)  to
select the  ``hot'' or ``cold''  solution from the line  profile fits.
The  mass  and  the  radius  of  the  white  dwarf  is  then  obtained
interpolating  the  determined $\Teff$  and  $\log  g$ in  the  tables
provided  by \citet{bergeronetal95-2}\footnote{Updated  tables can  be
  found                                                             at
  \emph{http://www.astro.umontreal.ca/$\sim$bergeron/CoolingModels/}}. However,
it has been shown that one-dimensional white dwarf model spectra such as
those used  in this work  yield overestimated $\log g$  (and therefore
mass)  values  for  white   dwarfs  of  effective  temperatures  below
$\sim$12\,000\,K    \citep[e.g.][]{koesteretal09-1,   tremblayetal11-1}.
Therefore, we apply the corrections of \citet{tremblayetal13-1}, which
are based  on three-dimensional  white dwarf model  spectra, to  our white
dwarf  parameter determinations.   Finally,  we use  the flux  scaling
factors  between the  observed residual  white dwarf  spectra and  the
best-fit white  dwarf models  to estimate  the white  dwarf distances,
which    are   equivalent    to    the   distances    to   our    DA/M
binaries\footnote{In  principle,  we  could  also  get  a
  distance estimate from the flux  scaling factor between the observed
  spectra and best-fit templates to  the secondary stars.  However, it
  has  been  shown that  $\sim$1/3  of  the secondary  star  distances
  obtained in  this way are  systematically larger than  the distances
  measured to the  white dwarf components, presumably  due to magnetic
  activity  \citep{rebassa-mansergasetal07-1}.   Therefore we  do  not
  take into account  the secondary star distances in  this work.}.  It
is  important to  mention that  the relative  flux calibration  of the
LAMOST  spectra  may  incorporate   systematic  uncertainties  in  the
distance determinations,  as the  flux scaling  factor depends  on the
absolute level  of the  flux-calibrated spectra,  and we  will discuss
this effect  in detail  below.  Table\,\ref{t-cat} provides  the white
dwarf stellar parameters of 65 DA/M binaries, and the M dwarf spectral
types of 104  DA/M binaries.  The LAMOST spectra of  the DA/M binaries
for which we could not obtain stellar parameters are either too noisy,
are dominated by  the flux from one of the  two stellar components, or
are subject  to a bad  flux calibration due  to lack of  suitable flux
standard stars. Because of these reasons we are able to obtain
  white dwarf stellar parameters and  spectral types of the companions
  for 27  and 40  of the  41 DA/M binaries  commonly observed  by both
  LAMOST and SDSS, respectively.

\begin{figure}
\centering
\includegraphics[width=\columnwidth]{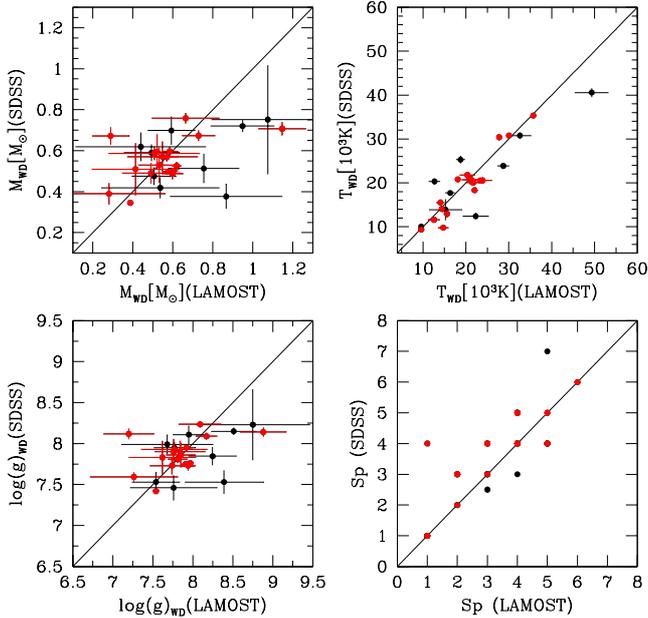}
\caption{Comparison  of the  white  dwarf mass  (top left),  effective
  temperature  (top   right),  surface  gravity  (bottom   left),  and
  secondary star spectral  type (bottom right) obtained for  27 (40 in
  the bottom  right panel) DA/M  binaries commonly observed  by LAMOST
  and SDSS.  In red are shown  systems for which their LAMOST and SDSS
  spectra have a S/N higher than 12.}
\label{fig:comp}
\end{figure}

We compare the white dwarf parameters of the 27 DA/M binaries
and the  spectral types of  the companions of the  40 systems
commonly observed by LAMOST and SDSS in Figure\,\ref{fig:comp}.  It is
important to emphasize here that  the stellar parameters obtained from
both the  LAMOST and  the SDSS  DA/M binary spectra  rely on  the same
decomposition/fitting routine. The  reliability of this method
  has     been     tested     by     \citet{rebassa-mansergasetal07-1,
    rebassa-mansergasetal10-1}, who  compared their  stellar parameter
  determinations from  SDSS spectra  to those obtained  by independent
  spectroscopic  parameter fitting  methods applied  to the  same SDSS
  spectra   \citep{raymondetal03-1,  vanetal05-1,   silvestrietal07-1,
    helleretal09-1}.   This  comparison   revealed  that  all  studies
  provide stellar  parameters that  are overall consistent  with those
  obtained    from    our     decomposition/fitting    routine    (see
  \citealt{rebassa-mansergasetal07-1,  rebassa-mansergasetal10-1}  for
  further  details).    Therefore,  directly  comparing   the  stellar
  parameters obtained here from LAMOST  spectra to those obtained from
  SDSS spectra  allows us  to test  in a direct  way how  reliable our
  measured     stellar     parameters     are.      Inspection     of
Figure\,\ref{fig:comp} reveals that the obtained spectral types are in
good agreement  (we find that  in $\sim$50\,per cent of the  cases the
spectral types are the same, and that only in two cases the difference
is of  more than  one spectral  subclass), and  that the  white dwarf
stellar parameters agree at the 1.5$\sigma$ level in $\sim$70\,per cent
of  the cases  (this  percentage  drops to  $\sim$40\,per cent if  we
consider the  white dwarf effective  temperatures; top right  panel of
Figure\,\ref{fig:comp}).  Because systematic uncertainties in the flux
calibration affect both  the SDSS and LAMOST spectra,  it is difficult
to assess  from which spectra  our routine provides the  most reliable
stellar  parameters.   However,  and  not surprisingly,  we  find  the
S/N  of the LAMOST and/or the  SDSS spectra to
be systematically  lower for the  cases with the  larger discrepancies
between  the  measured  values.   Indeed, if  we  consider  only  DA/M
binaries for  which the S/N  of both the  LAMOST and SDSS  spectra are
higher than  12 (red solid  dots in Figure\,\ref{fig:comp}),  then the
vast majority of cases associated with the larger discrepancies in the
measured parameters  disappear.\footnote{In order  to ensure  a direct
  comparison we estimate the S/N of the LAMOST and SDSS spectra in the
  same way as $\frac{1}{n} \sum  \frac{f}{ef}$, where $f$ and $ef$ are
  the flux and flux error respectively, and $n$ is the total number of
  data  points.}   We  therefore  only consider  further  the  stellar
parameters  of 48  LAMOST DA/M  binaries for  which the  S/N of  their
spectra is higher than 12.

\begin{figure}
\centering
\includegraphics[angle=-90, width=\columnwidth]{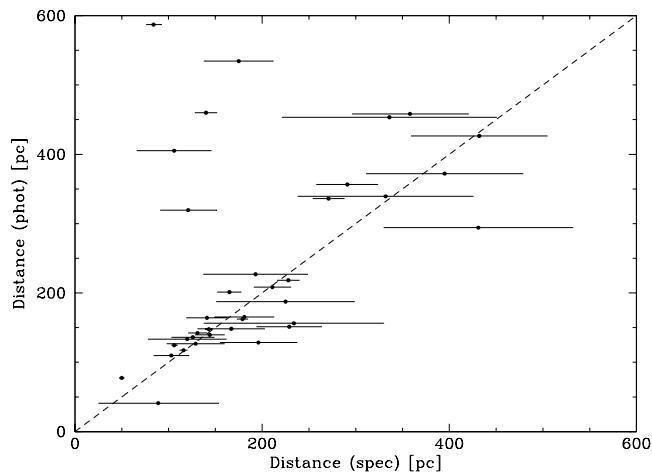}
\caption{Photometric  vs. spectroscopic  distances to  36 LAMOST  DA/M
  binaries with  available SDSS  DR\,8 $ugriz$ magnitudes  and LAMOST
  spectra of S/N ratio above 12.}
\label{f-distance}
\end{figure}

In order to investigate the effect of the relative flux calibration of
the  LAMOST spectra  in  our distance  determinations  we followed  an
independent  approach for  estimating the  distances and  compared the
results. We  restricted this exercise to  the 48 systems in  which the
LAMOST spectra are of S/N higher than 12.  We interpolated the $\Teff$
and $\log  g$ values  determined for  each white  dwarf in  these DA/M
binaries in  the tables  by \citet{bergeronetal95-2} and  obtained the
white dwarf  absolute magnitudes $UGRIZ$.  We  then gathered available
DR\,8 $ugriz$  magnitudes for  36 of our  selected targets  and simply
derive their  distances from their  distance modulus.  Given  that the
apparent magnitudes correspond  to those of the binary  system and the
absolute magnitudes are  calculated for the white  dwarf components we
considered  only  the distance  obtained  from  the $u$-band  distance
modulus in  each case, as  the M dwarf  contribution in this  range is
generally  low.   We  compare  the  distances  obtained  in  this  way
(hereafter photometric  distances) to the distances  obtained from the
flux-scaling   factors   (hereafter    spectroscopic   distances)   in
Figure\,\ref{f-distance}. Inspection  of the  figure reveals  that the
distances are  in relatively good  agreement except for five  cases in
which  the spectroscopic  distances  are considerably  lower than  the
photometric distances.   In three of  these particular cases  the DA/M
binary  spectra are  dominated by  the contribution  of the  secondary
stars  (contributing also  with a  significant amount  of flux  in the
$u$-band), therefore their photometric  distances cannot be considered
as  reliable.   This  exercise  demonstrates that  the  relative  flux
calibration of LAMOST spectra does not seem to affect dramatically our
spectroscopic distance determinations and  we hereafter adopt these as
the  distances to  our DA/M  binaries.  These  values are  included in
Table\,\ref{t-cat}.  The fact that  two independent distance estimates
broadly agree for  the vast majority of the  considered cases provides
indirect  support for  the  white dwarf  parameters  derived from  the
LAMOST spectra being reliable, as both distance determinations rely to
some extent on the accuracy of these parameters.

\section{Characterization of the LAMOST WDMS binary population}
\label{s-comparison}

\begin{figure*}
\centering
\includegraphics[angle=-90,width=0.9\textwidth]{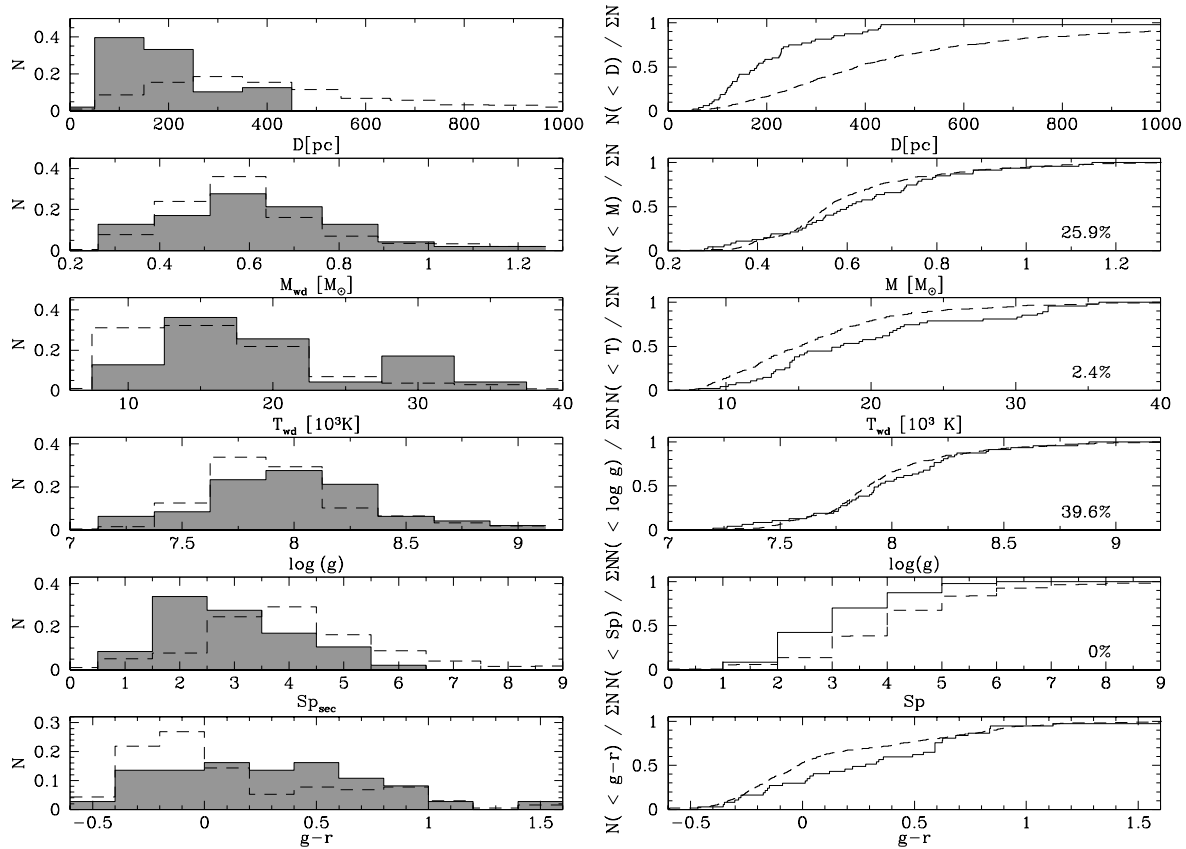}
\caption{From top to bottom:  normalised distributions (solid line for
  LAMOST  DA/M  binaries,  dashed  line for  SDSS  DA/M  binaries)  of
  distance,  white  dwarf  mass,  effective  temperature  and  surface
  gravity,  spectral type  of  the companion  star  and $g-r$  color.
  Kolmogorov-Smirnov  probabilities   resulting  from   comparing  the
  stellar parameter determinations from  both samples are indicated in
  the right panels.}
\label{fig:histo}
\end{figure*}

In this section  we study the intrinsic properties of  the LAMOST DA/M
binary population  and investigate  whether this  new sample  helps in
overcoming the selection effects  incorporated by the target selection
criteria  of  SDSS, which  is  biased  against the  identification  of
systems containing cool white dwarfs and early-type companions.  We do
this  analysing  and  comparing  the stellar  parameter  and  distance
distributions of  the LAMOST  and SDSS  DA/M binary  populations. However, we note
that the 251 SDSS DA/M binaries observed as part of SEGUE were
not considered here, as this population  was the result of a dedicated
search of  systems containing  systematically cooler white  dwarfs and
earlier-type    companions   \citep{rebassa-mansergasetal12-1}.     In
addition, we only consider in this  exercise systems for which the S/N
of  their  spectra  are  higher than  12  (Section\,\ref{s-param}),  a
restriction we applied both to LAMOST and SDSS DA/M binary spectra.

We begin  showing the  distance distribution of  LAMOST and  SDSS DA/M
binaries  in  the  top  left  panel  of  Figure\,\ref{fig:histo}.   To
facilitate the comparison both distributions have been normalised.  It
becomes  obvious  that, in  general,  the  LAMOST DA/M  binary  sample
contains  systems  that  are  considerably nearer  to  us  than  those
observed  by SDSS,  i.e.  LAMOST  observes  a smaller  and more  local
volume of DA/M binaries.  This is further illustrated in the top right
panel  of  Figure\,\ref{fig:histo},  where  we  provide  the  distance
cumulative  distributions.   As  we   will  show  below,  this
  difference  is  likely  a  consequence of  the  different  observing
  conditions of the two surveys, i.e.  LAMOST DA/M binaries need to be
  nearer to us  for their LAMOST spectra reaching high  enough S/N for
  determining          reliable          stellar          parameters.
\citet{rebassa-mansergasetal10-1} showed that selection effects affect
differently the  observed population depending on  the distance.  Thus
for  example, the  fraction of  systems containing  cool white  dwarfs
and/or  late-type   companions  increases  for   decreasing  distances
\citep[][see their  Figure\,16]{rebassa-mansergasetal10-1}. In
  the same  way, DA/M binaries  containing hotter white  dwarfs and/or
  earlier type  companions dominate  at larger distances.   Given that
  (1) due  to selection  effects the  DA/M binary  observed population
  depends on the distance considered and  (2) the LAMOST and SDSS DA/M
  binary distance distributions are  considerably different, all SDSS
DA/M  binaries located  at distances  larger than  450\,pc (the  upper
limit  distance of  the LAMOST  DA/M binary  sample) were  not further
considered.  The total number of LAMOST and SDSS DA/M binaries used in
this analysis was thus 48 and 398, respectively.

The middle four left panels  of Figure\,\ref{fig:histo} show the white
dwarf  parameter  and  the  secondary star  spectral  type  normalised
distributions of our 48 LAMOST DA/M binaries.  For comparison, we also
show the  normalised stellar parameter  distributions of the  398 SDSS
DA/M binaries.  In order to compare quantitatively the two populations
we applied Kolmogorov-Smirnov (KS) tests  to the white dwarf parameter
cumulative distributions, and apply a  $\chi^2$ test to the cumulative
spectral type distributions (right panels of Figure\,\ref{fig:histo}).
We obtained a  $\chi^2$ probability of 0 and KS  probabilities of 0.02
(effective temperature), 0.39 (surface gravity) and 0.26 (mass).

We  mentioned  above  that,  due  to  selection  effects,  the
  observed DA/M binary population  depends on the distance considered.
  Thus,   \citet{rebassa-mansergasetal10-1}  demonstrated   that  the
fraction   of   SDSS   DA/M  binaries   containing   relatively   cool
($\la$10\,000\,K) white  dwarfs and/or late type  ($\ga$M4) companions
increases  for decreasing  distances.  This  is simple  to understand:
cool  white dwarfs  (late companions  stars)  become too  faint to  be
detected at large distances, however  are easily accessible at shorter
distances  such as  those  considered  here ($\sim$50--450\,pc).   This
effect  can clearly  be  seen in  the spectral  type  and white  dwarf
effective   temperature   distributions    of   SDSS   DA/M   binaries
(Figure\,\ref{fig:histo}),  which  contain  a  large  number  of  such
systems.  In contrast, the LAMOST  DA/M binary population is dominated
by hotter  white dwarfs and  earlier type  companions.  This may  be a
consequence of the different target selection criteria used by the two
surveys.  However, we  have shown that $g-r \ga 0.7$  for $\sim$40\,per cent
  of  LAMOST   and  $\sim$30\,per cent  of   SDSS  DA/M  binaries
(Section\,\ref{s-selection}),  typical colors  of systems  containing
cooler      white      dwarfs      and      later-type      companions
\citep{rebassa-mansergasetal10-1}.   Therefore, the  fraction of  such
systems in LAMOST should actually be larger than in SDSS, which is not
the case (see  also the bottom panels  of Figure\,\ref{fig:histo}). It
is important to keep in mind  that we are considering in this analysis
only LAMOST and SDSS DA/M binaries with spectra of S/N higher than 12,
a  condition that  seems  to exclude  a large  number  of LAMOST  DA/M
binaries containing  cool white  dwarfs and late-type  companions.  We
explain this as followed.

The  average seeing  conditions of  Xinglong Observing  Station, where
LAMOST  is  located,  are  considerably worse  than  in  Apache  Point
Observatory, the  observational site  of SDSS. Hence  larger exposures
are needed  for LAMOST  to reach the  same S/N as  SDSS would  if both
surveys observed the same system.  However, the exposure times used by
LAMOST  and  SDSS are  roughly  similar  (typical exposure  times  are
30--90\,min,  depending on  the  considered plate),  hence the  LAMOST
spectrum of  a given DA/M binary  would generally have a  lower S/N as
compared to an  SDSS spectrum of the same system.   Thus DA/M binaries
containing cool white dwarfs and  late type companions that are easily
observed by  SDSS at $\sim$50--450\,pc  are generally either  too faint
for LAMOST to  be observed, or LAMOST observed them  and the resulting
spectra are of  too low S/N for obtaining  reliable stellar parameters
(i.e. they  are excluded  from this analysis).   This reflects  in the
peak at M3--4 in the spectral  type distribution of SDSS DA/M binaries
shifting  to M2--3  for  LAMOST  DA/M binaries,  and  in  the lack  of
late-type ($>$M6)  companions.  In the  same way this  effect explains
both the scarcity  of white dwarfs cooler  than $\Teff\la$\,15\,000\,K and
the  pronounced peak  at  $\Teff\sim$\,15\,000--20\,000\,K in  the  white
dwarf  effective temperature  distribution  of  LAMOST DA/M  binaries.
This naturally explains the null  $\chi^2$ and very low ($\sim$2.4\,per cent)
  KS  probability  obtained  when comparing  the  secondary  star
spectral  type  and  white   dwarf  effective  temperature  cumulative
distributions of both populations (Figure\,\ref{fig:histo}).

\begin{table*}
\centering
\caption{\label{t-rvs}   \Lines{Na}{I}{8183.27,  8194.81}   absorption
  doublet and  H$\alpha$ emission  radial velocities measured  for the
  LAMOST DA/M  binaries. Heliocentric  Julian dates and  the telescope
  used  for obtaining  the spectra  (either LAMOST  or SDSS)  are also
  indicated. The complete table can be found in the electronic version
  of the paper. ``-'' indicates that no radial velocity is available.}
\setlength{\tabcolsep}{1.4ex}
\begin{tabular}{ccccccc}
\hline
\hline
 Name                &      HJD       &    RV(NaI) & err & RV(H$\alpha$) & err & telescope \\
                     &   [days]       &   [km/s]  &      & [km/s]       & \\
\hline
J002952.22+434205.8   &     2456266.01670 &   31.90 &  11.58 &   57.23 &  11.69 &  LAMOST \\
J002952.22+434205.8   &     2456266.03649 &   35.86 &  12.11 &   50.08 &  11.06 &  LAMOST \\
J004432.63+402635.2   &     2455886.07410 &  -34.50 &  15.34 &       - &      - &  LAMOST \\
J005221.55+440335.4   &     2456206.22816 &    8.05 &  14.00 &       - &      - &  LAMOST \\
J005221.55+440335.4   &     2456206.24664 &   21.90 &  15.67 &       - &      - &  LAMOST \\
J005221.55+440335.4   &     2456207.14976 &       - &      - &   32.44 &  11.19 &  LAMOST \\
J005221.55+440335.4   &     2456207.16534 &   19.19 &  15.16 &   35.74 &  12.10 &  LAMOST \\
J005636.91+383332.6   &     2455913.98067 &   -0.61 &  11.73 &       - &      - &  LAMOST \\
J005636.91+383332.6   &     2455914.00629 &    3.49 &  11.90 &       - &      - &  LAMOST \\
J005601.44+400718.4   &     2455886.04888 &   -3.34 &  16.11 &       - &      - &  LAMOST \\
\hline
\end{tabular}
\end{table*}

Based on the above arguments, LAMOST  DA/M binaries may be expected to
also contain systematically lower-mass white dwarfs, as low-mass white
dwarfs have larger radii and are  therefore more luminous (for a given
effective  temperature).  However,  even though  the fraction  of such
white dwarfs is slightly larger  among LAMOST DA/M binaries, the white
dwarf  surface gravity  and mass  distributions are  similar for  both
LAMOST and  SDSS systems  (Figure\,\ref{fig:histo}).  This  is further
supported   by  the   KS  probabilities   derived  by   comparing  the
corresponding  cumulative distributions  ($\sim$26\,per cent for  the
mass, and  $\sim$40\,per cent  for the surface  gravity distributions).
At  a given  temperature, a  0.4\,\Msun white  dwarf is  $\sim$1.3 times
larger   than  a   0.6\,\Msun  white   dwarf  \citep{bergeronetal95-2,
  rebassa-mansergasetal11-1},  i.e.   it  has a  luminosity  $\sim$1.7
times larger. Conversely,  for a given surface gravity (or  mass if we
keep the radius fixed), the luminosity  of a 15\,000\,K white dwarf is
$\sim$5 times  larger than a $\sim$10\,000\,K  white dwarf. Therefore,
the contribution of  the effective temperature to the  luminosity of a
white dwarf is generally much more  important than that of its radius.
This  fact  may explain  why  the  population of  LAMOST  DA/M
  binaries  is  not  systematically dominated  by  systems  containing
  low-mass white dwarfs.

We conclude that LAMOST DA/M binaries are statistically different than
SDSS  DA/M  binaries.  This  is  not  a consequence  of  the
different  target   selection  criteria   of  the  two   surveys,  but
likely due  to the observing site  conditions being generally
worse at Xinglong Observing Station  than at Apache point Observatory,
where the  LAMOST and the  SDSS telescopes are  located, respectively.
This  effect naturally  explains  why the  LAMOST DA/M  binary
population  is  found between  $\sim$50--450\,pc  and  is dominated  by
systems  containing  hot  white   dwarfs  and  early-type  companions.
However, it is important to keep in mind that our conclusion is 
mainly based on the low KS and $\chi^2$  probabilities obtained when 
comparing the LAMOST and  SDSS white dwarf effective temperature and 
secondary spectral type distributions, respectively. Low KS probabilities 
may also be obtained as a consequence of different datasets leading to 
different  quality determinations,  a possibility we cannot rule out 
here as the  uncertainties in  the white dwarf stellar  parameters derived  
from LAMOST spectra are generally larger (see Figure\,\ref{fig:comp}).

\section{Identification of close LAMOST WDMS binaries}

In this section  we make use of all available  LAMOST and SDSS spectra
of our LAMOST DA/M binaries to analyse the fraction of close binaries,
i.e.  PCEBs,  in our sample.   We detect  close binaries via  a radial
velocity  (RV)  analysis based  on  multi-epoch  spectroscopy.  If  we
detect significant (more than 3$\sigma$)  RV variation we consider the
system as a close binary. If we do not detect RV variation we consider
the  system as  a wide  binary  candidate. This is because the  probability
exists for  the RVs  to sample the  same orbital phase  of a  PCEB, in
which   case  we   would   not  detect   RV  variation;   furthermore,
low-inclination and  long orbital period  PCEBs are more  difficult to
identify.

We  measured  the  RVs  fitting  the  \Lines{Na}{I}{8183.27,  8194.81}
absorption doublet and/or the H$\alpha$ emission line of each spectrum
following  the method  described by  \citet{rebassa-mansergasetal08-1,
  renetal13-1}.  This routine  fits the \Lines{Na}{I}{8183.27,8194.81}
absorption   doublet   with  a   second   order   polynomial  plus   a
double-Gaussian line  profile of  fixed separation, and  the H$\alpha$
emission line  with a second  order polynomial plus  a single-Gaussian
line profile.  We were able to obtain  at least one RV value for 63 of
our LAMOST DA/M binaries, 41 of which have available RVs measured from
the H$\alpha$ emission  line, 46 from the \Ion{Na}{I}  doublet, and 22
with RVs  from both  the H$\alpha$ emission  line and  the \Ion{Na}{I}
doublet.   However, only  for  42  systems we  did obtain  at least  two
\Ion{Na}{I}  and/or  two  H$\alpha$  emission RVs,  suitable  for  the
detection   of    close   binaries.   The   RVs    are   provided   in
Table\,\ref{t-rvs}.

We  detect  more  than  3$\sigma$ \Ion{Na}{I}  RV  variation  in  four
systems: J0759+3219, J0958-0200, J1054+5122 and J1649+1412. J0759+3219
and   J1054+5122    were   previously    identified   as    PCEBs   by
\citet{rebassa-mansergasetal11-1, schreiberetal10-1}.   J0958-0200 and
J1649+1412 are new  PCEBs displaying RV variation  from LAMOST spectra
obtained during the same  night.  Therefore, J0958-0200 and J1649+1412
are two  excellent candidates  for being  short orbital  period PCEBs.
Furthermore, we detect  H$\alpha$ emission RV variation  in 8 systems:
the four  mentioned above  displaying also significant  \Ion{Na}{I} RV
variation,  plus   J0757+3230,  J0819+0604,   J1125+2905,  J1347+2707.
However, the RVs  measured from the \Ion{Na}{I} doublet  do not differ
significantly for J0757+3230  and J1347+2707. It is  worth noting that
\citet{rebassa-mansergasetal08-1}  already  pointed   out  that  using
H$\alpha$ emission  RVs for identifying  PCEBs may not be  adequate in
some cases, therefore we wish to emphasize that our H$\alpha$ emission
RVs should be considered  with caution. In particular, further \Ion{Na}{I}
RVs are  needed to confirm  that the PCEBs identified here solely  by H$\alpha$
emission RV variation  are indeed close binaries.   This is definitely
not the case for J0757+3230 and J1347+2707.

From the above exercise we find that 4/30, i.e.  $\sim$13\,per cent, of
our LAMOST DA/M binaries with  more than two available \Ion{Na}{I} RVs
are PCEBs.  If  we consider only systems with  spectra taken separated
by  at least  one night  \citep[otherwise  we are  biased against  the
  detection  of   PCEBs  of   orbital  periods  longer   than  
$\sim$1\,day;][]{schreiberetal10-1}  the close  binary fraction  increases to
2/6,  i.e.  $\sim$33\,per cent.   Although we  suffer from  low number
statistics, this  value is  similar to the  PCEB fraction  found among
SDSS DA/M binaries  \citep[$\sim$21--24\,per cent,][]{nebotetal11-1} and
agrees  broadly with  the $\sim$25\,per cent  predicted by  population
synthesis studies \citep[e.g.][]{willems+kolb04-1}.   However, we need
to take  into account that the  PCEB fraction depends on  the spectral
type  of the  secondary star,  as angular  momentum loss  via magnetic
braking is  much more efficient  in decreasing the  orbital separation
when the secondary star is partially convective ($\sim$M0--3) than when
it  is  fully  convective   ($\sim$M4--9).   Therefore,  DA/M  binaries
containing $\sim$M0--3  companions become semi-detached faster  and the
PCEB    fraction     is    lower     at    these     spectral    types
\citep{politano+weiler06-1,  schreiberetal10-1}.   The secondary  star
spectral types of the four systems  we have identified as PCEBs are M2
(J0759+3219), M3 (J0958-0200 and J1054+5122) and M4 (J1649+1412).  The
measured SDSS  PCEB fraction  at these  spectral types  increases from
$\sim$20\,per cent   at   M2   to   $\sim$40\,per cent   at   M4
\citep{schreiberetal10-1}.    Given  that   the  LAMOST   DA/M  binary
population  is dominated  by systems  containing M2--3  secondary stars
(Figure\,\ref{fig:histo}), the  real PCEB  fraction among  LAMOST DA/M
binaries  should not  be  too  far from  the  value  obtained here  of
$\sim$33\,per cent.

\section{Summary and  Conclusions}

We have  developed an efficient,  fast and  novel method based  on the
wavelet  transform  to  identify   white  dwarf-main  sequence  (WDMS)
binaries containing  a DA  white dwarf and  a M-dwarf  companion (DA/M
binaries) within  the data release 1  of LAMOST.  We find  121 of such
systems, 80 of  which are new discoveries, and  estimate our catalogue
to contain $\sim$90\,per cent of  all DA/M binaries observed by LAMOST.
We expect  the spectra of  the $\sim$10\,per cent of DA/M  binaries we
missed to lack  from important data points in the  areas which must be
used for the wavelet transform, to be clearly dominated by the flux of
one of the two stars, or to be too noisy for a clear identification of
the spectral features of both stellar components.

We have  applied a  decomposition/fitting routine  to the  LAMOST DA/M
binary  spectra  to  measure  the white  dwarf  parameters  (effective
temperature,  surface gravity  and  mass), obtain  the secondary  star
spectral types,  and estimate the  distances to our systems.   We find
that reliable stellar parameters and  distances can be obtained for 48
DA/M binaries having LAMOST spectra of S/N above 12.
The stellar parameter  and distance distributions are  used to analyse
the properties  of the  LAMOST DA/M binary  sample and  to investigate
whether or  not this  population is  statistically different  from the
SDSS DA/M binary population, which  is biased towards the detection of
systems containing  cool white  dwarfs and early-type  companions.  We
find  that   the  LAMOST   and  SDSS   DA/M  binary   populations  are
different. However, this is not  due to the different target selection
criteria of both surveys, but  likely a simple consequence of
the different observing site conditions,  which are generally worse at
Xinglong Observing  Station than  at Apache Point  Observatory.  Thus,
the population of LAMOST DA/M binaries is concentrated at 50--450\,pc
and is dominated by systems containing hot white dwarfs and early-type
companions.   Even though  LAMOST  also fails  at identifying  systems
containing cool white dwarfs, the  LAMOST DA/M binary sample dominated
by  early-type companions  is  an important  addition  to the  current
sample of known spectroscopic  SDSS WDMS binaries.  Forthcoming LAMOST
data releases will allow us to identify more of such systems and to thus
build-up the small current sample of LAMOST (DR\,1) DA/M binaries.

We have  also identified four  DA/M binaries in our  sample displaying
significant  radial velocity  variation,  i.e.  PCEBs.   Two of  these
systems  were previously  identified as  close binaries  by SDSS,  the
other two are new PCEBs of expected short orbital periods. We find the
fraction of PCEBs among LAMOST DA/M binaries to be $\sim$33\,per cent.

\vspace{0.5cm}

{\it Acknowledgments.}--- This  work is supported by  the National Key
Basic Research of China 2014CB845700  and the National Natural Science
Foundation of China under  grant numbers 11390371, 11103031, 11233004,
and 61273248.   Guoshoujing Telescope (LAMOST) is  a National
Major   Scientific   Project  built   by   the   Chinese  Academy   of
Sciences. Funding  for the project  has been provided by  the National
Development and Reform  Commission. LAMOST is operated  and managed by
the National Astronomical Observatories,  Chinese Academy of Sciences.
The     LAMOST      Data     Release      One     Web      site     is
\emph{http://data.lamost.org/dr1/}.  We   thank  Detlev   Koester  for
providing  us with  his grid  of white  dwarf model  spectra, and  the
anonymous  referee for  his/her comments  and suggestions  that helped
improving  the  quality  of  the paper.   ARM  acknowledges  financial
support  from  the Postdoctoral  Science  Foundation  of China  (grant
2013M530470)  and  from  the  Research Fund  for  International  Young
Scientists by the National Natural  Science Foundation of China (grant
11350110496).


\begin{thebibliography}{57}
\expandafter\ifx\csname natexlab\endcsname\relax\def\natexlab#1{#1}\fi

\bibitem[{{Adelman-McCarthy} {et~al.}(2008){Adelman-McCarthy}, {Ag{\"u}eros},
  {Allam}, {Allende Prieto}, {Anderson}, {Anderson}, {Annis}, {Bahcall},
  {Bailer-Jones}, {Baldry}, {Barentine}, {Bassett}, {Becker}, {Beers}, {Bell},
  {Berlind}, {Bernardi}, {Blanton}, {Bochanski}, {Boroski}, {Brinchmann},
  {Brinkmann}, {Brunner}, {Budav{\'a}ri}, {Carliles}, {Carr}, {Castander},
  {Cinabro}, {Cool}, {Covey}, {Csabai}, {Cunha}, {Davenport}, {Dilday}, {Doi},
  {Eisenstein}, {Evans}, {Fan}, {Finkbeiner}, {Friedman}, {Frieman},
  {Fukugita}, {G{\"a}nsicke}, {Gates}, {Gillespie}, {Glazebrook}, {Gray},
  {Grebel}, {Gunn}, {Gurbani}, {Hall}, {Harding}, {Harvanek}, {Hawley},
  {Hayes}, {Heckman}, {Hendry}, {Hindsley}, {Hirata}, {Hogan}, {Hogg}, {Hyde},
  {Ichikawa}, {Ivezi{\'c}}, {Jester}, {Johnson}, {Jorgensen}, {Juri{\'c}},
  {Kent}, {Kessler}, {Kleinman}, {Knapp}, {Kron}, {Krzesinski}, {Kuropatkin},
  {Lamb}, {Lampeitl}, {Lebedeva}, {Lee}, {Leger}, {L{\'e}pine}, {Lima}, {Lin},
  {Long}, {Loomis}, {Loveday}, {Lupton}, {Malanushenko}, {Malanushenko},
  {Mandelbaum}, {Margon}, {Marriner}, {Mart{\'{\i}}nez-Delgado}, {Matsubara},
  {McGehee}, {McKay}, {Meiksin}, {Morrison}, {Munn}, {Nakajima}, {Neilsen},
  {Newberg}, {Nichol}, {Nicinski}, {Nieto-Santisteban}, {Nitta}, {Okamura},
  {Owen}, {Oyaizu}, {Padmanabhan}, {Pan}, {Park}, {Peoples}, {Pier}, {Pope},
  {Purger}, {Raddick}, {Re Fiorentin}, {Richards}, {Richmond}, {Riess}, {Rix},
  {Rockosi}, {Sako}, {Schlegel}, {Schneider}, {Schreiber}, {Schwope}, {Seljak},
  {Sesar}, {Sheldon}, {Shimasaku}, {Sivarani}, {Smith}, {Snedden}, {Steinmetz},
  {Strauss}, {SubbaRao}, {Suto}, {Szalay}, {Szapudi}, {Szkody}, {Tegmark},
  {Thakar}, {Tremonti}, {Tucker}, {Uomoto}, {Vanden Berk}, {Vandenberg},
  {Vidrih}, {Vogeley}, {Voges}, {Vogt}, {Wadadekar}, {Weinberg}, {West},
  {White}, {Wilhite}, {Yanny}, {Yocum}, {York}, {Zehavi}, \&
  {Zucker}}]{adelman-mccarthyetal08-1}
{Adelman-McCarthy}, J.~K., {Ag{\"u}eros}, M.~A., {Allam}, S.~S., {et~al.} 2008, ApJS,
  175, 297

\bibitem[{{Bergeron} {et~al.}(1995){Bergeron}, {Wesemael}, \& {Beauchamp}}]{bergeronetal95-2}
{Bergeron}, P., {Wesemael}, F., \& {Beauchamp}, A. 1995, \pasp, 107, 1047

\bibitem[{{Camacho} {et~al.}(2014){Camacho}, {Torres}, {Garc{\'{\i}}a-Berro},
  {Zorotovic}, {Schreiber}, {Rebassa-Mansergas}, {Nebot G{\'o}mez-Mor{\'a}n},
  \& {G{\"a}nsicke}}]{camachoetal14-1}
{Camacho}, J., {Torres}, S., {Garc{\'{\i}}a-Berro}, E., {et~al.} 2014, ArXiv
  e-prints

\bibitem[{{Carlin} {et~al.}(2012){Carlin}, {L{\'e}pine}, {Newberg}, {Deng}, {Beers}, {Chen}, {Christlieb}, {Fu}, {Gao}, {Grillmair}, {Guhathakurta}, {Han}, {Hou}, {Lee}, {Li}, {Liu}, {Liu}, {Pan}, {Sellwood}, {Wang}, {Yang},  {Yanny}, {Zhang}, {Zheng}, \& {Zhu}}]{carlinetal12-1}
{Carlin}, J.~L., {L{\'e}pine}, S., {Newberg}, H.~J., {et~al.} 2012, Research in Astronomy and Astrophysics, 12, 755

\bibitem[{{Chen} {et~al.}(2012){Chen}, {Hou}, {Yu}, {Liu}, {Deng}, {Newberg}, {Carlin}, {Yang}, {Zhang}, {Shen}, {Zhang}, {Chen}, {Chen}, {Christlieb}, {Han}, {Lee}, {Liu}, {Pan}, {Shi}, {Wang}, \& {Zhu}}]{chenetal12-1}
{Chen}, L., {Hou}, J.-L., {Yu}, J.-C., {et~al.} 2012, Research in Astronomy and Astrophysics, 12, 805

\bibitem[{{Chui}(1992)}]{chui92-1}
{Chui}, C.~K. 1992, {Wavelets: A tutorial in theory and applications}

\bibitem[{{Cui} {et~al.}(2012){Cui}, {Zhao}, {Chu}, {Li}, {Li}, {Zhang}, {Su}, {Yao}, {Wang}, {Xing}, {Li}, {Zhu}, {Wang}, {Gu}, {Luo}, {Xu}, {Zhang}, {Liu}, {Zhang}, {Yang}, {Cao}, {Chen}, {Chen}, {Chen}, {Chen}, {Chu}, {Feng},
  {Gong}, {Hou}, {Hu}, {Hu}, {Hu}, {Jia}, {Jiang}, {Jiang}, {Jiang}, {Jin}, {Li}, {Li}, {Li}, {Liu}, {Liu}, {Lu}, {Mao}, {Men}, {Qi}, {Qi}, {Shi}, {Tang}, {Tao}, {Wang}, {Wang}, {Wang}, {Wang}, {Wang}, {Wang}, {Wang}, {Wang}, {Wang}, {Wang}, {Wang}, {Wang}, {Xu}, {Xu}, {Yang}, {Yu}, {Yuan}, {Yuan}, {Zhai}, {Zhang}, {Zhang}, {Zhang}, {Zhao}, {Zhou}, {Zhou}, {Zhu}, \& {Zou}}]{cuietal12-1}
{Cui}, X.-Q., {Zhao}, Y.-H., {Chu}, Y.-Q., {et~al.} 2012, Research in Astronomy and Astrophysics, 12, 1197

\bibitem[{{De Marco} {et~al.}(2011){De Marco}, {Passy}, {Moe}, {Herwig}, {Mac
  Low}, \& {Paxton}}]{demarcoetal11-1}
{De Marco}, O., {Passy}, J.-C., {Moe}, M., {et~al.} 2011, \mnras, 411, 2277

\bibitem[{{Davis} {et~al.}(2010){Davis}, {Kolb}, \& {Willems}}]{davisetal10-1}
{Davis}, P.~J., {Kolb}, U., \& {Willems}, B. 2010, \mnras, 403, 179

\bibitem[{{Deng} {et~al.}(2012){Deng}, {Newberg}, {Liu}, {Carlin}, {Beers}, {Chen}, {Chen}, {Christlieb}, {Grillmair}, {Guhathakurta}, {Han}, {Hou}, {Lee}, {L{\'e}pine}, {Li}, {Liu}, {Pan}, {Sellwood}, {Wang}, {Wang}, {Yang}, {Yanny}, {Zhang}, {Zhang}, {Zheng}, \& {Zhu}}]{dengetal12-1}
{Deng}, L.-C., {Newberg}, H.~J., {Liu}, C., {et~al.} 2012, Research in Astronomy and Astrophysics, 12, 735

\bibitem[{{Farihi} {et~al.}(2010){Farihi}, {Hoard}, \& {Wachter}}]{farihietal10-1}
{Farihi}, J., {Hoard}, D.~W., \& {Wachter}, S. 2010, \apjs, 190, 275

\bibitem[{{Ferrario}(2012)}]{ferrario12-1}
{Ferrario}, L. 2012, \mnras, 426, 2500

\bibitem[{{Heller} {et~al.}(2009){Heller}, {Homeier}, {Dreizler}, \& {{\O}stensen}}]{helleretal09-1}
{Heller}, R., {Homeier}, D., {Dreizler}, S., \& {{\O}stensen}, R. 2009, \aap, 496, 191

\bibitem[{{Iben} \& {Livio}(1993)}]{iben+livio93-1}
{Iben}, I.~J. \& {Livio}, M. 1993, \pasp, 105, 1373

\bibitem[{{Koester}(2010)}]{koester10-1}
{Koester}, D. 2010, \memsai, 81, 921

\bibitem[{{Koester} {et~al.}(2009){Koester}, {Kepler}, {Kleinman}, \&
  {Nitta}}]{koesteretal09-1}
{Koester}, D., {Kepler}, S.~O., {Kleinman}, S.~J., \& {Nitta}, A. 2009, Journal
  of Physics Conference Series, 172, 012006

\bibitem[{{Li}(2012)}]{Li2012}
{Li}, X.-R. 2012, Spectroscopy and Spectral Analysis, 30, 94

\bibitem[{{Liu} {et~al.}(2014){Liu}, {Yuan}, {Huo}, {Deng}, {Hou}, {Zhao}, {Zhao}, {Shi}, {Luo}, {Xiang}, {Zhang}, {Huang}, \& {Zhang}}]{liuetal14-1}
{Liu}, X.-W., {Yuan}, H.-B., {Huo}, Z.-Y., {et~al.} 2014, IAU Symposium, 298, 310 

\bibitem[{{Luo} {et~al.}(2012){Luo}, {Zhang}, {Zhao}, {Zhao}, {Cui}, {Li}, {Chu}, {Shi}, {Wang}, {Zhang}, {Bai}, {Chen}, {Wang}, {Guo}, {Chen}, {Du}, {Kong}, {Lei}, {Li}, {Song}, {Wu}, {Zhang}, {Zhou}, {Zuo}, {Du}, {He}, {Hou}, {Dong}, {Li}, {Li}, {Li}, {Song}, {Tian}, {Wang}, {Wu}, {Yang}, {Yuan}, {Cao}, {Chen}, {Chen}, {Chen}, {Chu}, {Feng}, {Gong}, {Gu}, {Hou}, {Huo}, {Hu}, {Hu}, {Hu}, {Jia}, {Jiang}, {Jiang}, {Jiang}, {Jin}, {Li}, {Li}, {Li}, {Li}, {Li}, {Liu}, {Liu}, {Liu}, {Lu}, {Lu}, {Luo}, {Mao}, {Men}, {Ni}, {Qi}, {Qi}, {Shi}, {Su}, {Sun}, {Su}, {Tang}, {Tao}, {Tu}, {Wang}, {Wang}, {Wang}, {Wang}, {Wang}, {Wang}, {Wang}, {Wang}, {Wang}, {Wang}, {Wang}, {Wang}, {Wang}, {Wang}, {Wei}, {Xue}, {Xing}, {Xu}, {Xu}, {Xu}, {Yang}, {Yang}, {Yao}, {Yu}, {Yuan}, {Zhai}, {Zhang}, {Zhang}, {Zhang}, {Zhang}, {Zhang}, {Zhang}, {Zhao}, {Zhou}, {Zhu}, {Zhu}, \& {Zou}}]{luoetal12-1}
{Luo}, A.-L., {Zhang}, H.-T., {Zhao}, Y.-H., {et~al.} 2012, Research in Astronomy and Astrophysics, 12, 1243

\bibitem[{{Marsh} {et~al.}(2014){Marsh}, {Parsons}, {Bours}, {Littlefair},
  {Copperwheat}, {Dhillon}, {Breedt}, {Caceres}, \&
  {Schreiber}}]{marshetal14-1}
{Marsh}, T.~R., {Parsons}, S.~G., {Bours}, M.~C.~P., {et~al.} 2014, \mnras,
  437, 475

\bibitem[{{Morgan} {et~al.}(2012){Morgan}, {West}, {Garc{\'e}s}, {Catal{\'a}n}, {Dhital}, {Fuchs}, \& {Silvestri}}]{morganetal12-1}
{Morgan}, D.~P., {West}, A.~A., {Garc{\'e}s}, A., {et~al.} 2012, \aj, 144, 93

\bibitem[{{Nebot G{\'o}mez-Mor{\'a}n} {et~al.}(2011){Nebot G{\'o}mez-Mor{\'a}n}, {G{\"a}nsicke}, {Schreiber}, {Rebassa-Mansergas}, {Schwope}, {Southworth}, {Aungwerojwit}, {Bothe}, {Davis}, {Kolb}, {M{\"u}ller}, {Papadaki}, {Pyrzas}, {Rabitz}, {Rodr{\'{\i}}guez-Gil}, {Schmidtobreick}, {Schwarz}, {Tappert}, {Toloza}, {Vogel}, \&  {Zorotovic}}]{nebotetal11-1}
{Nebot G{\'o}mez-Mor{\'a}n}, A., {G{\"a}nsicke}, B.~T., {Schreiber}, M.~R., {et~al.} 2011, \aap, 536, A43

\bibitem[{{Nebot G{\'o}mez-Mor{\'a}n} {et~al.}(2009){Nebot G{\'o}mez-Mor{\'a}n}, {Schwope}, {Schreiber}, {G{\"a}nsicke}, {Pyrzas}, {Schwarz}, {Southworth}, {Kohnert}, {Vogel}, {Krumpe}, \& {Rodr{\'{\i}}guez-Gil}}]{nebotetal09-1}
{Nebot G{\'o}mez-Mor{\'a}n}, A., {Schwope}, A.~D., {Schreiber}, M.~R., {et~al.} 2009, \aap, 495, 561

\bibitem[{{Parsons} {et~al.}(2014){Parsons}, {Marsh}, {Bours}, {Littlefair},
  {Copperwheat}, {Dhillon}, {Breedt}, {Caceres}, \&
  {Schreiber}}]{parsonsetal14-1}
{Parsons}, S.~G., {Marsh}, T.~R., {Bours}, M.~C.~P., {et~al.} 2014, \mnras,
  438, L91

\bibitem[{{Parsons} {et~al.}(2010){Parsons}, {Marsh}, {Copperwheat}, {Dhillon},
  {Littlefair}, {G{\"a}nsicke}, \& {Hickman}}]{parsonsetal10-1}
{Parsons}, S.~G., {Marsh}, T.~R., {Copperwheat}, C.~M., {et~al.} 2010, \mnras,
  402, 2591

\bibitem[{{Parsons} {et~al.}(2012{\natexlab{a}}){Parsons}, {Marsh},
  {G{\"a}nsicke}, {Dhillon}, {Copperwheat}, {Littlefair}, {Pyrzas}, {Drake},
  {Koester}, {Schreiber}, \& {Rebassa-Mansergas}}]{parsonsetal12-2}
{Parsons}, S.~G., {Marsh}, T.~R., {G{\"a}nsicke}, B.~T., {et~al.}
  2012{\natexlab{a}}, \mnras, 419, 304

\bibitem[{{Parsons} {et~al.}(2012{\natexlab{b}}){Parsons}, {Marsh},
  {G{\"a}nsicke}, {Rebassa-Mansergas}, {Dhillon}, {Littlefair}, {Copperwheat},
  {Hickman}, {Burleigh}, {Kerry}, {Koester}, {Nebot G{\'o}mez-Mor{\'a}n},
  {Pyrzas}, {Savoury}, {Schreiber}, {Schmidtobreick}, {Schwope}, {Steele}, \&
  {Tappert}}]{parsonsetal12-1}
{Parsons}, S.~G., {Marsh}, T.~R., {G{\"a}nsicke}, B.~T., {et~al.}
  2012{\natexlab{b}}, \mnras, 420, 3281

\bibitem[{{Politano} \& {Weiler}(2006)}]{politano+weiler06-1}
{Politano}, M. \& {Weiler}, K.~P. 2006, \apjl, 641, L137

\bibitem[{{Pyrzas} {et~al.}(2012){Pyrzas}, {G{\"a}nsicke}, {Brady}, {Parsons}, {Marsh}, {Koester}, {Breedt}, {Copperwheat}, {Nebot G{\'o}mez-Mor{\'a}n}, {Rebassa-Mansergas}, {Schreiber}, \& {Zorotovic}}]{pyrzasetal12-1}
{Pyrzas}, S., {G{\"a}nsicke}, B.~T., {Brady}, S., {et~al.} 2012, \mnras, 419, 817

\bibitem[Raymond et al.(2003)]{raymondetal03-1} 
Raymond, S.~N., Szkody, P., Hawley, S.~L., et al.\ 2003, \aj, 125, 2621 

\bibitem[{{Rebassa-Mansergas} {et~al.}(2007){Rebassa-Mansergas}, {G{\"a}nsicke}, {Rodr{\'{\i}}guez-Gil}, {Schreiber}, \& {Koester}}]{rebassa-mansergasetal07-1}
{Rebassa-Mansergas}, A., {G{\"a}nsicke}, B.~T., {Rodr{\'{\i}}guez-Gil}, P., {Schreiber}, M.~R., \& {Koester}, D. 2007, \mnras, 382, 1377

\bibitem[Rebassa-Mansergas et al.(2008)]{rebassa-mansergasetal08-1} 
Rebassa-Mansergas, A., G{\"a}nsicke, B.~T., Schreiber, M.~R., et al.\ 2008, \mnras, 390, 1635 

\bibitem[{{Rebassa-Mansergas} {et~al.}(2010){Rebassa-Mansergas}, {G{\"a}nsicke}, {Schreiber}, {Koester}, \&  {Rodr{\'{\i}}guez-Gil}}]{rebassa-mansergasetal10-1}
{Rebassa-Mansergas}, A., {G{\"a}nsicke}, B.~T., {Schreiber}, M.~R., {Koester}, D., \& {Rodr{\'{\i}}guez-Gil}, P. 2010, \mnras, 402, 620

\bibitem[{{Rebassa-Mansergas} {et~al.}(2011){Rebassa-Mansergas}, {Nebot G{\'o}mez-Mor{\'a}n}, {Schreiber}, {Girven}, \& {G{\"a}nsicke}}]{rebassa-mansergasetal11-1}
{Rebassa-Mansergas}, A., {Nebot G{\'o}mez-Mor{\'a}n}, A., {Schreiber}, M.~R., {Girven}, J., \& {G{\"a}nsicke}, B.~T. 2011, \mnras, 413, 1121
  
\bibitem[Rebassa-Mansergas et al.(2012{\natexlab{a}})]{rebassa-mansergasetal12-1} 
Rebassa-Mansergas, A., Nebot G{\'o}mez-Mor{\'a}n, A., Schreiber, M.~R., et al.\ 2012, \mnras, 419, 806 

\bibitem[{{Rebassa-Mansergas} {et~al.}(2012{\natexlab{b}}){Rebassa-Mansergas}, {Zorotovic}, {Schreiber}, {G{\"a}nsicke}, {Southworth}, {Nebot G{\'o}mez-Mor{\'a}n}, {Tappert}, {Koester}, {Pyrzas}, {Papadaki}, {Schmidtobreick}, {Schwope}, \& {Toloza}}]{rebassa-mansergasetal12-2}
{Rebassa-Mansergas}, A., {Zorotovic}, M., {Schreiber}, M.~R., {et~al.} 2012, \mnras, 423, 320

\bibitem[{{Rebassa-Mansergas} {et~al.}(2013{\natexlab{b}}){Rebassa-Mansergas}, {Schreiber}, \& {G{\"a}nsicke}}]{rebassa-mansergasetal13-1}
{Rebassa-Mansergas}, A., {Schreiber}, M.~R., \& {G{\"a}nsicke}, B.~T. 2013{\natexlab{b}}, \mnras, 429, 3570

\bibitem[{{Rebassa-Mansergas} {et~al.}(2013{\natexlab{a}}){Rebassa-Mansergas}, {Agurto-Gangas}, {Schreiber}, {G{\"a}nsicke}, \& {Koester}}]{rebassa-mansergasetal13-2}
{Rebassa-Mansergas}, A., {Agurto-Gangas}, C., {Schreiber}, M.~R., {G{\"a}nsicke}, B.~T., \& {Koester}, D. 2013{\natexlab{a}}, \mnras, 433, 3398

\bibitem[{{Ren} {et~al.}(2013){Ren}, {Luo}, {Li}, {Wei}, {Zhao}, {Zhao}, {Song}, \& {Zhao}}]{renetal13-1}
{Ren}, J., {Luo}, A., {Li}, Y., {et~al.} 2013, \aj, 146, 82

\bibitem[{{Richards} {et~al.}(2002){Richards}, {Fan}, {Newberg}, {Strauss},
  {Vanden Berk}, {Schneider}, {Yanny}, {Boucher}, {Burles}, {Frieman}, {Gunn},
  {Hall}, {Ivezi{\'c}}, {Kent}, {Loveday}, {Lupton}, {Rockosi}, {Schlegel},
  {Stoughton}, {SubbaRao}, \& {York}}]{richardsetal02-1}
{Richards}, G.~T., {Fan}, X., {Newberg}, H.~J., {et~al.} 2002, \aj, 123, 2945

\bibitem[{{Schreiber} {et~al.}(2010){Schreiber}, {G{\"a}nsicke}, {Rebassa-Mansergas}, {Nebot Gomez-Moran}, {Southworth}, {Schwope}, {M{\"u}ller}, {Papadaki}, {Pyrzas}, {Rabitz}, {Rodr{\'{\i}}guez-Gil}, {Schmidtobreick}, {Schwarz}, {Tappert}, {Toloza}, {Vogel}, \& {Zorotovic}}]{schreiberetal10-1}
{Schreiber}, M.~R., {G{\"a}nsicke}, B.~T., {Rebassa-Mansergas}, A., {et~al.} 2010, \aap, 513, L7+

\bibitem[Silvestri et al.(2007)]{silvestrietal07-1} 
Silvestri, N.~M., Lemagie, M.~P., Hawley, S.~L., et al.\ 2007, \aj, 134, 741 

\bibitem[{{Song} {et~al.}(2012){Song}, {Luo}, {Comte}, {Bai}, {Zhang}, {Du}, {Zhang}, {Chen}, {Zuo}, \& {Zhao}}]{songetal12-1}
{Song}, Y.-H., {Luo}, A.-L., {Comte}, G., {et~al.} 2012, Research in Astronomy and Astrophysics, 12, 453

\bibitem[Smol{\v c}i{\'c} et al.(2004)]{smolcicetal04-1} 
Smol{\v c}i{\'c}, V., Ivezi{\'c}, {\v Z}., Knapp, G.~R., et al.\ 2004, \apjl, 615, L141

\bibitem[{{Starck} {et~al.}(1997){Starck}, {Siebenmorgen}, \& {Gredel}}]{starcketal97-1}
{Starck}, J.-L., {Siebenmorgen}, R., \& {Gredel}, R. 1997, \apj, 482, 1011

\bibitem[{{Strauss} {et~al.}(2002){Strauss}, {Weinberg}, {Lupton}, {Narayanan},
  {Annis}, {Bernardi}, {Blanton}, {Burles}, {Connolly}, {Dalcanton}, {Doi},
  {Eisenstein}, {Frieman}, {Fukugita}, {Gunn}, {Ivezi{\' c}}, {Kent}, {Kim},
  {Knapp}, {Kron}, {Munn}, {Newberg}, {Nichol}, {Okamura}, {Quinn}, {Richmond},
  {Schlegel}, {Shimasaku}, {SubbaRao}, {Szalay}, {Vanden Berk}, {Vogeley},
  {Yanny}, {Yasuda}, {York}, \& {Zehavi}}]{straussetal02-1}
{Strauss}, M.~A., {Weinberg}, D.~H., {Lupton}, R.~H., {et~al.} 2002, \aj, 124, 1810

\bibitem[{{Tremblay} {et~al.}(2011){Tremblay}, {Ludwig}, {Steffen}, {Bergeron},
  \& {Freytag}}]{tremblayetal11-1}
{Tremblay}, P.-E., {Ludwig}, H.-G., {Steffen}, M., {Bergeron}, P., \&
  {Freytag}, B. 2011, \aap, 531, L19

\bibitem[{{Tremblay} {et~al.}(2013){Tremblay}, {Ludwig}, {Steffen}, \& {Freytag}}]{tremblayetal13-1}
{Tremblay}, P.-E., {Ludwig}, H.-G., {Steffen}, M., \& {Freytag}, B. 2013, \aap, 559, A104

\bibitem[van den Besselaar et al.(2005)]{vanetal05-1} 
van den Besselaar, E.~J.~M., Roelofs, G.~H.~A., Nelemans, G.~A., Augusteijn, T., \& Groot, P.~J.\ 2005, \aap, 434, L13 

\bibitem[{{Webbink}(2008)}]{webbink07-1}
{Webbink}, R.~F. 2008, in Astrophysics and Space Science Library, Vol. 352, Astrophysics and Space Science Library, ed. {E.~F.~Milone, D.~A.~Leahy, \& D.~W.~Hobill}, 233--+

\bibitem[{{Wei} {et~al.}(2013){Wei}, {Luo}, {Li}, {Pan}, {Tu}, {Jiang}, {Kong}, {Shi}, {Yi}, {Wang}, {Liu}, \& {Zhao}}]{weietal13-1}
{Wei}, P., {Luo}, A., {Li}, Y., {et~al.} 2013, \mnras, 431, 1800

\bibitem[{{Willems} \& {Kolb}(2004)}]{willems+kolb04-1}
{Willems}, B. \& {Kolb}, U. 2004, \aap, 419, 1057

\bibitem[Xiang et al.(2014)]{xiangetal14-1}
Xiang M.~S., Liu X.~W., Yuan H.~B., Huo Z.~Y., Huang Y., Zheng Y., Chen B.~Q. 2014, \mnras, in preparation

\bibitem[Yanny et al.(2009)]{yannyetal09-1} 
Yanny, B., Rockosi, C., Newberg, H.~J., et al.\ 2009, \aj, 137, 4377 

\bibitem[{{York} {et~al.}(2000){York}, {Adelman}, {Anderson}, {Anderson}, {Annis}, {Bahcall}, {Bakken}, {Barkhouser}, {Bastian}, {Berman}, {Boroski}, {Bracker}, {Briegel}, {Briggs}, {Brinkmann}, {Brunner}, {Burles}, {Carey}, {Carr}, {Castander}, {Chen}, {Colestock}, {Connolly}, {Crocker}, {Csabai}, {Czarapata}, {Davis}, {Doi}, {Dombeck}, {Eisenstein}, {Ellman}, {Elms}, {Evans}, {Fan}, {Federwitz}, {Fiscelli}, {Friedman}, {Frieman}, {Fukugita},  {Gillespie}, {Gunn}, {Gurbani}, {de Haas}, {Haldeman}, {Harris}, {Hayes}, {Heckman}, {Hennessy}, {Hindsley}, {Holm}, {Holmgren}, {Huang}, {Hull}, {Husby}, {Ichikawa}, {Ichikawa}, {Ivezi{\'c}}, {Kent}, {Kim}, {Kinney}, {Klaene},{Kleinman}, {Kleinman}, {Knapp}, {Korienek}, {Kron}, {Kunszt}, {Lamb}, {Lee}, {Leger}, {Limmongkol}, {Lindenmeyer}, {Long}, {Loomis}, {Loveday}, {Lucinio}, {Lupton}, {MacKinnon}, {Mannery}, {Mantsch}, {Margon}, {McGehee}, {McKay}, {Meiksin}, {Merelli}, {Monet}, {Munn}, {Narayanan}, {Nash}, {Neilsen}, {Neswold}, {Newberg}, {Nichol}, {Nicinski}, {Nonino}, {Okada}, {Okamura}, {Ostriker}, {Owen}, {Pauls}, {Peoples}, {Peterson}, {Petravick}, {Pier}, {Pope}, {Pordes}, {Prosapio}, {Rechenmacher}, {Quinn}, {Richards}, {Richmond}, {Rivetta}, {Rockosi}, {Ruthmansdorfer}, {Sandford}, {Schlegel}, {Schneider}, {Sekiguchi}, {Sergey}, {Shimasaku}, {Siegmund}, {Smee}, {Smith}, {Snedden}, {Stone}, {Stoughton}, {Strauss}, {Stubbs}, {SubbaRao}, {Szalay}, {Szapudi}, {Szokoly}, {Thakar}, {Tremonti}, {Tucker}, {Uomoto}, {Vanden Berk}, {Vogeley}, {Waddell}, {Wang}, {Watanabe}, {Weinberg}, {Yanny}, \& {Yasuda}}]{yorketal00-1}
{York}, D.~G., {Adelman}, J., {Anderson}, J.~E., {et~al.} 2000, \aj, 120, 1579

\bibitem[{{Zhao} {et~al.}(2012){Zhao}, {Zhao}, {Chu}, {Jing}, \& {Deng}}]{zhaoetal12-1}
{Zhao}, G., {Zhao}, Y.-H., {Chu}, Y.-Q., {Jing}, Y.-P., \& {Deng}, L.-C. 2012, Research in Astronomy and Astrophysics, 12, 723

\bibitem[{{Zorotovic} \& {Schreiber}(2013)}]{zorotovic+schreiber13-1}
{Zorotovic}, M. \& {Schreiber}, M.~R. 2013, \aap, 549, A95

\bibitem[{{Zorotovic} {et~al.}(2010){Zorotovic}, {Schreiber}, {G{\"a}nsicke}, \& {Nebot G{\'o}mez-Mor{\'a}n}}]{zorotovicetal10-1}
{Zorotovic}, M., {Schreiber}, M.~R., {G{\"a}nsicke}, B.~T., \& {Nebot G{\'o}mez-Mor{\'a}n}, A. 2010, \aap, 520, A86

\end{thebibliography}

\end{document}